\documentclass[fleqn,usenatbib,useAMS]{mnras}

\usepackage{graphicx}	
\usepackage{xcolor}
\usepackage{amsmath}	
\usepackage{amssymb}	
\usepackage{multicol}        
\usepackage{bm}		
\usepackage{pdflscape}	
\usepackage[normalem]{ulem}
\usepackage{dblfloatfix}

\newcommand{\hi}{H\,{\sc i}}
\newcommand{\hii}{H\,{\sc ii}}
\newcommand{\arepo}{\textsc{Arepo}}

\newcommand{\ethos}{\textsc{Ethos}}
\newcommand{\ethostwo}{\textsc{sDAO}}
\newcommand{\ethosfour}{\textsc{Ethos-4}}

\usepackage[T1]{fontenc}
\usepackage{ae,aecompl}

\usepackage{newtxtext,newtxmath}

\title[Lyman-$\alpha$ forest and DAOs]{ETHOS -- an Effective Theory of Structure Formation: detecting dark matter interactions through the Lyman-$\alpha$ forest}

\author[S. Bose et al.]{Sownak Bose$^{1}$\thanks{Contact
    e-mail: \href{mailto:sownak.bose@cfa.harvard.edu}{sownak.bose@cfa.harvard.edu}},
  Mark Vogelsberger$^{2}$, Jes{\'u}s Zavala$^{3}$, Christoph
  Pfrommer$^{4}$,\newauthor Francis-Yan
  Cyr-Racine$^{5,6}$, Sebastian Bohr$^{3}$ and Torsten Bringmann$^{7}$
  \\
$^{1}$Harvard-Smithsonian Center for Astrophysics, 60 Garden Street, Cambridge, MA 02138, USA \\
$^{2}$Department of Physics, Kavli Institute for Astrophysics and Space Research, Massachusetts Institute of Technology, Cambridge, MA 02139, USA \\
$^{3}$Center for Astrophysics and Cosmology, Science Institute, University of Iceland, Dunhagi 5, 107 Reykjavik, Iceland \\
$^{4}$Leibniz-Institut f{\"u}r Astrophysik Potsdam, An der Sternwarte 16, 14482 Potsdam, Germany \\
$^{5}$Department of Physics, Harvard University, Cambridge, MA 02138, USA \\
$^{6}$Department of Physics and Astronomy, University of New Mexico, Albuquerque, NM 87131, USA \\
$^{7}$Department of Physics, University of Oslo, Box 1048, N-0371 Oslo, Norway}

\date{}

\pubyear{2018}

\begin{document}
\label{firstpage}
\pagerange{\pageref{firstpage}--\pageref{lastpage}}
\maketitle

\begin{abstract}
We perform a series of cosmological hydrodynamic simulations to investigate the effects of non-gravitational dark matter (DM) interactions on the intergalactic medium (IGM). In particular, we use the {\sc Ethos} framework \citep{CyrRacine2016,Vogelsberger2016} to compare statistics of the Lyman-$\alpha$ forest in cold dark matter (CDM) with an alternative model in which the DM couples strongly with a relativistic species in the early universe. These models are characterised by a cutoff in the linear power spectrum, followed by a series of `dark acoustic oscillations' (DAOs) on sub-dwarf scales. While the primordial cutoff delays the formation of the first galaxies, structure builds-up more rapidly in the interacting DM model compared to CDM. We show that although DAOs are quickly washed away in the non-linear clustering of DM at $z\lesssim10$, their signature can be imprinted prominently in the Lyman-$\alpha$ flux power spectrum at $z>5$. On scales larger than the cutoff ($k\sim0.08$ s/km for the specific model considered here), the relative difference to CDM is reminiscent of a warm dark matter (WDM) model with a similar initial cutoff; however, the redshift evolution on smaller scales is distinctly different. The appearance and disappearance of DAOs in the Lyman-$\alpha$ flux spectrum provides a powerful way to distinguish interacting DM models from WDM and, indeed, variations in the thermal history of the IGM that may also induce a small-scale cutoff. 
\end{abstract}




\section{Introduction}
\label{sect:intro}

In the standard picture of structure formation, the enigmatic dark matter (DM) particle is assumed to be a kinematically cold, collisionless and non-baryonic entity. This `cold' dark matter (CDM) model has predictive power, a feature that has been exploited over the past four decades in a rigorous campaign of numerical simulations that has established CDM as part of the standard cosmological paradigm. The great success of this model lies in the finding that the same theory that accounts for the temperature anisotropies in the cosmic microwave background at early times \citep{Spergel2003,Planck2016} has been similarly successful at reproducing the large-scale clustering of galaxies at present day \citep{Colless2001,Cole2005,Eisenstein2005,Zehavi2011}.

At the regime of dwarf galaxies, however, a number of ``small-scale challenges'' have been claimed to afflict the CDM paradigm. Chief amongst them are the so-called ``Missing Satellites'' \citep[e.g.][]{Klypin1999,Moore1999}, ``Too Big to Fail'' \citep{MBK2011}, cusp-core \citep[e.g.][]{Flores1994,deBlok2001} and plane of satellites problems \citep[e.g.][]{Ibata2014,Pawlowski2014}. For a thorough overview of this subject, we refer the reader to the recent review by \cite{Bullock2017}. While these issues have been used to motivate DM candidates beyond CDM, it should be cautioned that the small-scale problems are only firmly established with simulations that include the modelling of the CDM component only. These issues may in fact be resolved within the CDM paradigm once the impact of gas and stellar physics is better understood and fully taken into account. This has been permitted by the increasing sophistication of hydrodynamical simulations \citep[e.g.][]{Vogelsberger2014a,Vogelsberger2014b, Dubois2014,Schaye2015, Springel2018}, which self-consistently track the co-evolution of dark and baryonic matter. Although hydrodynamical simulations differ in detail, they have shown universally that the interaction between DM and baryons through processes associated with galaxy formation -- such as gas cooling, photoionisation and feedback -- change both the census of the galactic population \citep[e.g.][]{Kauffmann1993,Kim2017,Read2017}, as well as the internal structure of DM haloes \citep[e.g.][]{Pontzen2014,DiCintio2014,Sawala2016,Fitts2017} relative to simulations modelling only the DM component.

In fact, perhaps the greatest challenge to the CDM model, at least in terms of its appeal as a complete structure formation theory, is that despite intense efforts at discovering CDM-like particles, the search has been fruitless so far. Most of these efforts have focused on Weakly Interacting Massive Particles (WIMPs), which are one of the best-motivated CDM candidates in great part due to their potential connection with supersymmetry~\citep{Jungman:1995df}. After the successful discovery of the Higgs boson, it was hoped that the Large Hadron Collider (LHC) would find evidence for supersymmetry, giving credence to WIMPs and the CDM model, but thus far the LHC has failed to provide any evidence of this kind. WIMPs have also remained elusive to both direct \citep[][]{Aprile2018} and indirect  detection \citep[e.g.][]{Albert2017} experiments. Furthermore, promising observational anomalies that might be connected to DM, have either disappeared or explained with non-DM astrophysical sources, e.g., 
the excess of gamma rays at the Galactic Centre, which has been ascribed to the self-annihilation of WIMPs \citep[e.g.][]{Hooper2011,Daylan2016} may instead be explained by a population of unresolved millisecond pulsars \citep[e.g.][]{Bartels:2015aea,Lee2016,Fermi2017}, or an overdensity of stars in the Galactic bulge \citep{Macias2018}. 

As long as the DM remains undetected in the laboratory, it is worth considering well-motivated alternatives to CDM and their implications for structure formation. A popular alternative is {\it warm} dark matter \citep[WDM,][]{Bond1983,Colin2000,Bode2001}, in which the DM particles have a non-negligible velocity dispersion in the early universe. The resulting free-streaming of these particles suppresses density fluctuations below a characteristic scale determined by the rest mass of the particles and their thermal history; this delays the formation of the first structures and reduces the abundance of low-mass galaxies in the process \citep[e.g.][]{Zavala2009,Maccio2010,Lovell2012,Schneider2012,Kennedy2014,Bose2016,Bozek2018}. In the linear regime, the free-streaming of WDM particles is manifest as a nearly exponential cutoff in the linear power spectrum relative to CDM. In other well-motivated DM models, there may exist a coupling between the DM and a relativistic species (e.g. neutrinos or `dark' radiation) in the early universe. In these so-called `interacting' dark matter (iDM) models, the ensuing radiation pressure inhibits the growth of small-scale fluctuations and also results in a (collisional) cutoff in the linear power spectrum, but with a more complex behaviour than in WDM, exhibiting dark acoustic oscillations \citep[e.g.][]{Carlson1992,Boehm:2001hm,Ackerman:2008gi,Cyr-Racine:2013ab,Buckley2014,Boehm2014,Bringmann:2016ilk}. Another promising alternative is offered by self-interacting dark matter (SIDM) models, in which multiple scattering events between DM particles can significantly change the internal structure of DM haloes compared to CDM in the non-linear regime  \citep[e.g.][]{Spergel2000,Yoshida2000,Dave2001,Colin2002,Vogelsberger2012,Rocha2013,Zavala2013,Elbert2015,Vogelsberger2014c, Kaplinghat2016,Robertson2018,Vogelsberger2018}. 

While these different DM species come from a diverse range of particle physics models with vastly different production mechanisms, the resulting effect on structure formation is similar in many cases. This is particularly evident in the case of WDM and iDM, both of which suppress small-scale structure by inducing a cutoff in the linear power spectrum. In this work, we consider examples of these models within the generalised framework of structure formation \ethos{} \citep[][]{CyrRacine2016,Vogelsberger2016}, which addresses such degeneracies by providing a mapping between parameters associated with DM physics and parameters relevant for structure formation. The flexibility afforded by this formalism potentially allows the investigation of a general class of model parameters (cutoff scale, self-interaction cross-section, DM-radiation coupling etc.) and their impact on the formation of galaxies without needing to simulate every point in the allowed parameter space. Previous analyses connected to this programme have focused on the predictions of these models for the internal content of DM haloes \citep{Vogelsberger2016,Brinckmann2018,Sameie2018,Sokolenko:2018noz}, the diverse rotation curves of dwarf galaxies \citep{Creasey2017}, the tidal stripping of satellites in the Galactic halo \citep{Dooley2016} and the possibility of detecting these DM candidates through 
gravitational lensing \citep{Diaz2018}. Finally, the most recent investigation has considered the high redshift galaxy population and reionisation history in this general class of models \citep{Das:2017nub,Lovell:2017eec}. In this paper, we are particularly interested in the signatures of new DM phenomenology (see Section~\ref{sect:ethos_desc} for details) that may be imprinted in statistics of the Lyman-$\alpha$ forest.

The Lyman-$\alpha$ forest has proven to be a remarkably powerful probe of the nature of DM in the mildly non-linear regime. Measurements of the flux spectrum using observed QSO sightlines have been used repeatedly to infer the clustering of matter on these scales \citep[e.g.][]{Croft1998,Croft1999,McDonald2000,Palanque2013}. The flux spectrum is a particularly powerful probe of processes relating to early galaxy formation and the small-scale behaviour of DM particles by providing an insight on the matter power spectrum at relatively high redshift. In fact, it is now well-established that the {\it observed} flux spectrum displays a cutoff in power on scales smaller than $k \sim 0.03$ s/km, a feature that has been used to constrain the free-streaming properties of DM particles \cite[e.g.][]{Viel2005,Seljak2006,Viel2013,Baur2016,Irsic2017,Kobayashi2017,Murgia2018,Nori2018,Garzilli2018}. 

A cutoff in the flux spectrum towards small scales may, however, originate from purely baryonic processes \citep[e.g.][]{Zaldarriaga2001,Peeples2010,Rorai2013,Nasir2016}. The first effect, known as Jeans smoothing, is a result of increased gas pressure in the IGM due to boosted temperatures induced by the onset of reionisation. The degree of Jeans smoothing depends on both the integrated heat injection and exact timing of reionisation \citep[e.g.][]{Gnedin1998,Kulkarni2015,Onorbe2017}. A second effect is brought upon by random thermal motions of the gas, resulting in Doppler broadening of Lyman-$\alpha$ forest lines, further smoothing small-scale power in the flux spectrum. The degenerate behaviour of the thermal history of the IGM and the free-streaming properties of DM make it difficult to pinpoint the physical interpretation of the observed cutoff in the flux spectrum and may indeed relax current constraints on the rest mass of the WDM particle \citep{Garzilli2017}. Nevertheless, it is clear that the Lyman-$\alpha$ forest provides a unique probe into mildly non-linear scales at high redshift, which is the regime where most alternative DM models exhibit the strongest deviations from CDM. The main goal of this paper is to show that, if strong enough, non-gravitational features (dark acoustic oscillations) in the primordial power spectrum of iDM models can remain imprinted in the Lyman-$\alpha$ flux spectrum.

The layout of this paper is as follows. In Section~\ref{sect:ethos_desc}, we briefly describe the DM particle physics model considered in this work, highlighting its connection to the general \ethos{} framework. Section~\ref{sect:setup} describes the numerical setup used for this investigation, detailing our simulations and the analysis pipeline used to extract mock Lyman-$\alpha$ absorption spectra from them. Section~\ref{sect:results} presents our main findings. Finally, our conclusions are summarised in Section~\ref{sect:conclusions}.

\section{Dark Matter model}
\label{sect:ethos_desc}
 In this work, we study structure formation in DM theories in which early-universe interactions with a relativistic species \citep[see e.g.][]{Carlson1992,Boehm:2001hm,Ackerman:2008gi,Feng:2009mn,Aarssen:2012fx,Chu:2014lja,Buen-Abad:2015ova,Bringmann:2016ilk,Chacko:2016kgg} lead to a modified initial spectrum of density fluctuations as compared to standard CDM. The general phenomenology of such models is described in detail in \cite{CyrRacine2016} within the \ethos{} framework, while the nonlinear evolution of structure within these models was studied in \cite{Buckley2014} and \cite{Vogelsberger2016} \citep[see also][]{Boehm:2014vja,Schewtschenko:2014fca}. In these theories, the DM forms a fluid that is tightly-coupled to a relativistic species (e.g. neutrinos or dark radiation) at early times, much like the standard baryon-photon plasma before the epoch of recombination. Within this `dark' fluid, the large radiation pressure prohibits the growth of DM fluctuations and allows the propagation of acoustic waves to large cosmological distances \citep{Cyr-Racine:2013fsa}. Just like the more well-known baryon acoustic oscillations (BAOs), these \emph{dark acoustic oscillations} (DAOs) become imprinted on the spectrum of matter fluctuations at late times, providing us with a potential smoking gun for physical processes taking place in the early Universe. Due to the finite value of the coupling between DM and the relativistic species, the DAOs are usually damped on scales smaller than the radiation mean free path, in a process similar to standard Silk damping \citep{Silk:1967kq}.
 
 The resulting shape of the linear matter power spectrum is largely determined by how quickly the DM kinetically decouples from the radiation bath. Quantitatively, near the redshift of DM kinetic decoupling, $z_{\rm D}$, we have approximately:
\begin{equation}\label{eq:w_vs_s_DAO}
(\dot{\kappa}_\chi/\mathcal{H})|_{z \sim z_{\rm D}}\simeq (z/z_{\rm D})^n,    
\end{equation}
 where $\dot{\kappa}_\chi$ is the `drag opacity' or interaction rate between DM and the relativistic species, and $\mathcal{H}$ is the conformal Hubble expansion rate. Note that $\dot{\kappa}$ denotes a derivative with respect to conformal time. In general, a larger value for the power law index\footnote{We note that this power law index is the same as that used to classify DM models within the \ethos{} framework \citep{CyrRacine2016}.}, $n$, results in a greater number of undamped DAOs on the small-scale linear power spectrum. Once the non-linear evolution of density fluctuations is taken into account, models with low values of $n\lesssim4$ exhibit structure formation that is reminiscent of standard WDM models \citep{Vogelsberger2016,Murgia:2017lwo}. On the other hand, models characterised by a large value of the power law index $n\gtrsim6$ (which we hereafter refer to as ``strong'' DAO models; sDAO) have a structure formation history that is appreciably different from WDM, as first discussed in \cite{Buckley2014}. We note that the models used in \cite{Vogelsberger2016}, \cite{Lovell:2017eec}, and \cite{Das:2017nub} all have $n=4$, and thus fall in the former category. 
 
Since our aim is to study Lyman-$\alpha$ constraints on DM theories that have a structure formation history that is distinct from WDM, the present investigation focuses on sDAO models. In particular, we consider an atomic DM \citep{Kaplan:2009de,Kaplan:2011yj,Cyr-Racine:2013ab} in which DM is composed of two massive fermions that are oppositely charged under a new unbroken $U(1)$ dark gauge force. In this paradigm, the dark sector forms an ionised plasma at early times, until the temperature falls below the binding energy between the two oppositely charged particles, at which point neutral dark atoms form in a process reminiscent to cosmological hydrogen recombination. If this ``dark'' recombination occurs in or near thermal equilibrium, an extremely rapid kinematic decoupling epoch ensues due to the nearly exponential (Saha-like) nature of bound state formation in this case. This ensures that the power law index appearing in Eq.~\eqref{eq:w_vs_s_DAO} is large ($n=6$ for our sDAO model), resulting in a linear matter power spectrum composed of a significant number of undamped DAOs on small-scales.  
 
We choose parameters of this model such that the linear matter power spectrum of our sDAO model starts deviating from its CDM counterpart near a comoving wavenumber of $k\sim10h$ Mpc$^{-1}$, which is the scale where current observations of the Lyman-$\alpha$ spectrum become a powerful tool to discriminate different DM models. In the nomenclature of the \ethos{} framework, the \ethostwo{} model is defined by:
\begin{equation*}
    \left\{ n,a_n,\omega_{{\rm DR}}, \alpha_2, \alpha_{l\geq3} \right\} = \left\{ 6\,, 6\times10^{8} {\rm Mpc}^{-1}, 1.25\times10^{-8},9/10\,,1 \right\}
\end{equation*}
while the \ethosfour{} model\footnote{We note that the $a_4$ amplitudes given in \cite{Vogelsberger2016} should be divided by $h$ to yield the correct values.} is defined by:
\begin{equation*}
    \left\{ n,a_n,\omega_{{\rm DR}}, \alpha_{l\geq2} \right\} = \left\{ 4\,, 414\, {\rm Mpc}^{-1}, 1.35\times10^{-6},3/2 \right\}
\end{equation*}
where $n$ is the power law index defined in Eq.~(\ref{eq:w_vs_s_DAO}), $a_n$ is the normalisation of the drag opacity at redshift $z_{\rm D}=10^7$, $\omega_{\rm DR}\equiv\Omega_{{\rm DR}}h^2$ is the physical energy density in the dark radiation component in units of the critical density and $\alpha_l$ is a set of coefficients that defines the angular dependence of the DM-dark radiation scattering cross section. We refer the reader to Section II E of \cite{CyrRacine2016} for further details. The actual particle physics (i.e. Lagrangian) parameters of the \ethostwo{} model are listed in Appendix~\ref{app:model_params}.

Fig.~\ref{fig:inputPk} illustrates the power spectrum appropriate to this model. For comparison, we also display the matter power spectra for CDM, \ethosfour{}, as well as WDM thermal relics with mass $1.6$ and $3.3$ keV. The parameters for \ethosfour{} were especially selected to alleviate the Missing Satellites, Too Big To Fail and core-cusp ``problems'' in CDM (Section~\ref{sect:intro}) through DM physics alone \citep{Vogelsberger2016}. Unlike WDM, in which the cutoff continues indefinitely, the \ethos{} models show a resurgence of power on smaller scales due to the aforementioned DAOs. Note that because of these DAOs, \ethos{} models, in particular sDAO models, have increased small-scale power compared to WDM models with a cutoff at the same scale. We note that the \ethostwo{} model, which is our main focus in this paper, may already be strongly constrained by present observations. Our goal here is to investigate if small-scale DAOs may be at all detectable in the Lyman-$\alpha$ forest, rather than to construct a model that matches the data. For this reason, we opt to simulate an iDM scenario that maximises differences relative to CDM on scales large enough that they may be captured at moderate numerical expense.

\section{Numerical setup}
\label{sect:setup}
\subsection{Simulations and initial conditions}
\label{sect:sims}

\begin{figure}
    \centering
    \includegraphics[width=\columnwidth]{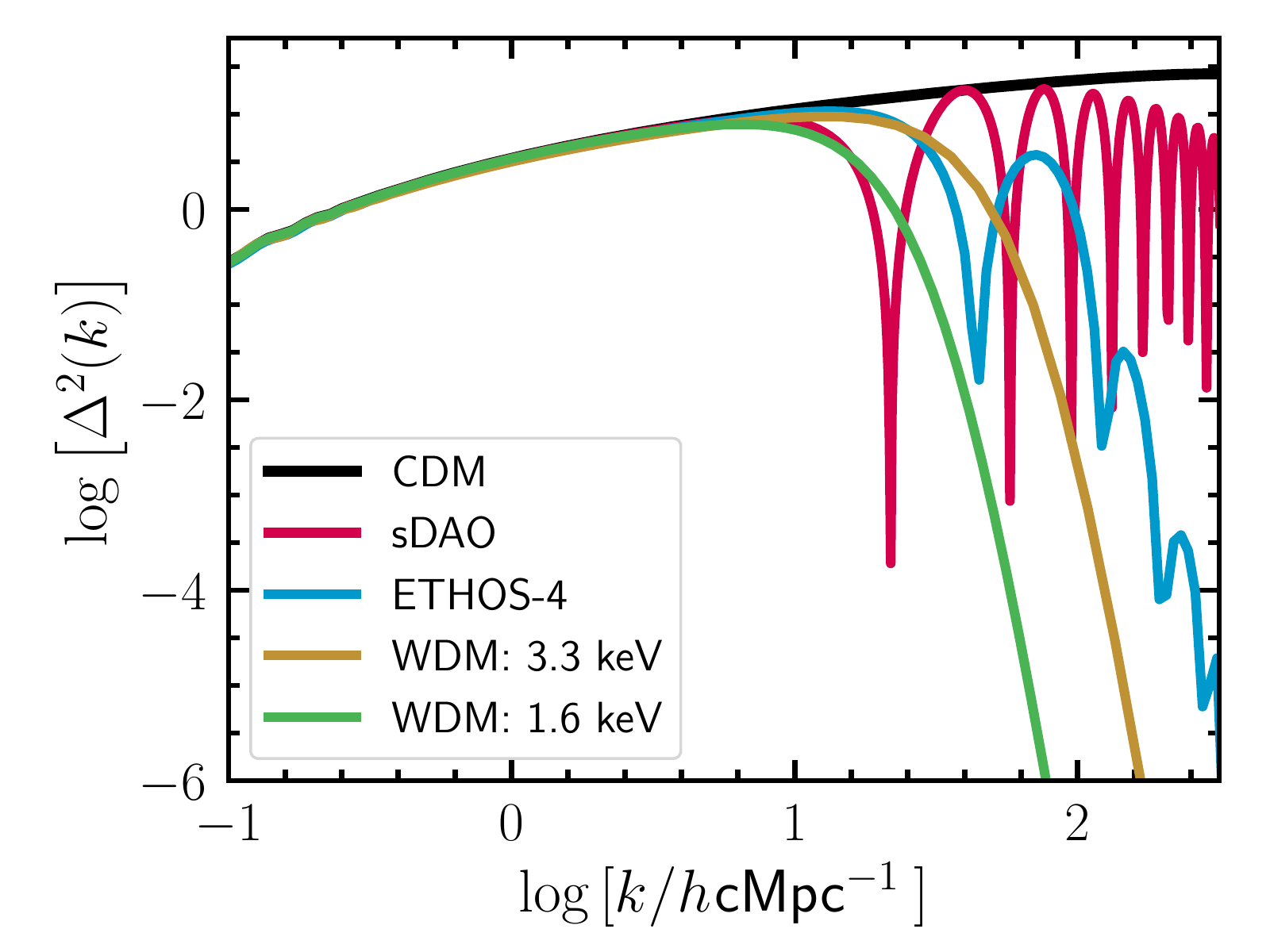}
    \caption{Dimensionless power spectra $\left[\Delta^2 (k) = k^3 P(k)\right]$ for the CDM (black) and \ethostwo{} (red) models used in this work. For comparison, we also show the power spectra for the less extreme \ethosfour{} model (blue; see \citealt{Vogelsberger2016}) which exhibits a deviation from CDM at a scale comparable to that of a 3.3 keV thermal relic WDM particle (in yellow). On the other hand, the cutoff scale for the \ethostwo{} model is closer to that of a 1.6 keV thermal relic (green). Furthermore, the amplitude of the dark acoustic oscillations (DAOs) in the \ethostwo{} model is considerably larger than in \ethosfour{}.}
    \label{fig:inputPk}
\end{figure}

The simulations we present in this work make use of the cosmological simulation code, \arepo{} \citep{Springel2010}. \arepo{} employs a hybrid tree/particle-mesh scheme to solve for gravitational interactions of DM particles, and a moving, unstructured Voronoi mesh to solve equations of hydrodynamics. The moving mesh is adaptive in nature, resolving fluids in regions of high density with many more cells of a smaller size than in low density environments. \arepo{} has been augmented with a comprehensive model for galaxy formation \citep[][]{Weinberger2017,Pillepich2018} which we use here. In addition, \cite{Vogelsberger2016} presents an updated version of \arepo{} which, in addition to the galaxy formation models mentioned above, also incorporates elastic, isotropic self-interactions of DM particles, while allowing for arbitrary velocity-dependent interaction cross-sections (using an algorithm adapted from the original described in detail in \citealt{Vogelsberger2012}). While the self-scatterings of DM particles have a pronounced impact on the internal structure of haloes at late-times, their influence on the IGM at high redshifts will be sub-dominant to that induced by the cutoff in the power spectrum; we have therefore turned off self-interactions in the simulations.

Our high resolution simulations follow the evolution of $2\times512^3$ DM and gas particles from $z=127$ to $z=0$ in a periodic box of (comoving) size $29.6$ cMpc (20 $h^{-1}$cMpc), resulting in an effective DM particle mass of $6.41\times10^6\,{\rm M}_\odot$. An individual gas cell has a target mass of $1.01\times10^6\,{\rm M}_\odot$. This target gas mass also corresponds to the typical mass of a stellar macro-particle representing a stellar population. We enforce that the mass of all cells is within a factor of two of the target mass by explicitly refining and de-refining the mesh cells. The comoving softening length for DM particles is set to 1.19 kpc, while the (adaptive) softening applied to a gas cell is set to a comoving minimum value of 185 pc. To check for convergence, we also run a second set of simulations a factor of two lower in resolution. 

We use the fiducial IllustrisTNG galaxy formation model \citep{Weinberger2017,Pillepich2018} with one change. Namely, we have turned off the magnetohydrodynamics solver as it is not relevant for the analysis presented here. As in the fiducial TNG model, each of our simulations is set up with a time-dependent, spatially uniform ionising background as described in the model by \cite{CFG2009}. The TNG model is built upon the original Illustris galaxy formation model described in \cite{Vogelsberger2013}.

Initial conditions for all simulations were generated using the {\sc music} code \citep{Hahn2011}, assuming cosmological parameters derived from \cite{Planck2016}: $\Omega_0 = 0.311$ (total matter density), $\Omega_{\rm b} = 0.049$ (baryon density), $\Omega_\Lambda = 0.689$ (dark energy density), $H_0 = 67.5$ kms$^{-1}$Mpc$^{-1}$ (Hubble parameter) and $\sigma_8 = 0.815$ (linear rms density fluctuation in a sphere of radius 8 $h^{-1}$ Mpc at $z=0$). The dimensionless linear power spectra used to generate initial conditions are shown in Fig.~\ref{fig:inputPk}. While the CDM power spectrum exhibits power on all scales, the two \ethos{} models cutoff at $\log [ k/h$cMpc$^{-1}] \approx 1$. 
In this paper we will be concerned with the \ethostwo{} model, in which the model parameters have been adjusted to amplify the effect of DAOs, as explained in the previous section. Our goal is to investigate the extent to which the characteristics of DAOs in the \ethos{} models can be probed using the Lyman-$\alpha$ forest. To put our results in context, we have also run simulations of the \ethosfour{} and 1.6 keV WDM models at our default resolution. The choice of a 1.6 keV thermal relic is motivated by the fact that the free-streaming scale in this model is identical to the cutoff in \ethostwo{}; this helps disentangle small-scale differences induced by the acoustic oscillations from those that are caused by a primordial cutoff. The simulations are analysed to perform mock Lyman-$\alpha$ observations using the procedure that we describe in the following subsection. Finally, we note that simulations that resolve the primordial power spectrum cutoff are plagued with artificial fragmentation of filaments that condense into `spurious' haloes \citep[e.g.][]{Wang2007,Lovell2014}. These objects are seeded by discreteness of the particle set rather than a true gravitational instability, and must hence be excluded from the analysis. This is a well-known problem in WDM simulations, but is less severe in the \ethos{} models which have added small-scale power in the form of DAOs \citep[][see also Fig.~\ref{fig:hmfCompare}]{Buckley2014}. This is especially true at high redshift, which is the regime of interest in this paper. As such, we do not perform any extra steps to classify these objects in the simulations we have run. 

\subsection{Creating Lyman-$\alpha$ mock absorption spectra}
\label{sect:lya}

From the outputs of each simulation, we generate synthetic absorption spectra using the methodology outlined in \cite{Altay2013}. In short, at each output time, we select 1024 randomly-selected skewers\footnote{We have checked explicitly that our results are converged for this choice for the number of sightlines (see Fig.~\ref{fig:NlosCompare}).} oriented parallel to a coordinate axis of the box. Gas cell properties are interpolated onto locations along each skewer using a smoothing kernel; we follow \cite{Altay2013} and employ a truncated Gaussian kernel, $G_t (r,\sigma)$, which is defined as:

\begin{figure*}
    \centering
    \includegraphics[width=\textwidth,trim = {0 0.8in 0 0}, clip]{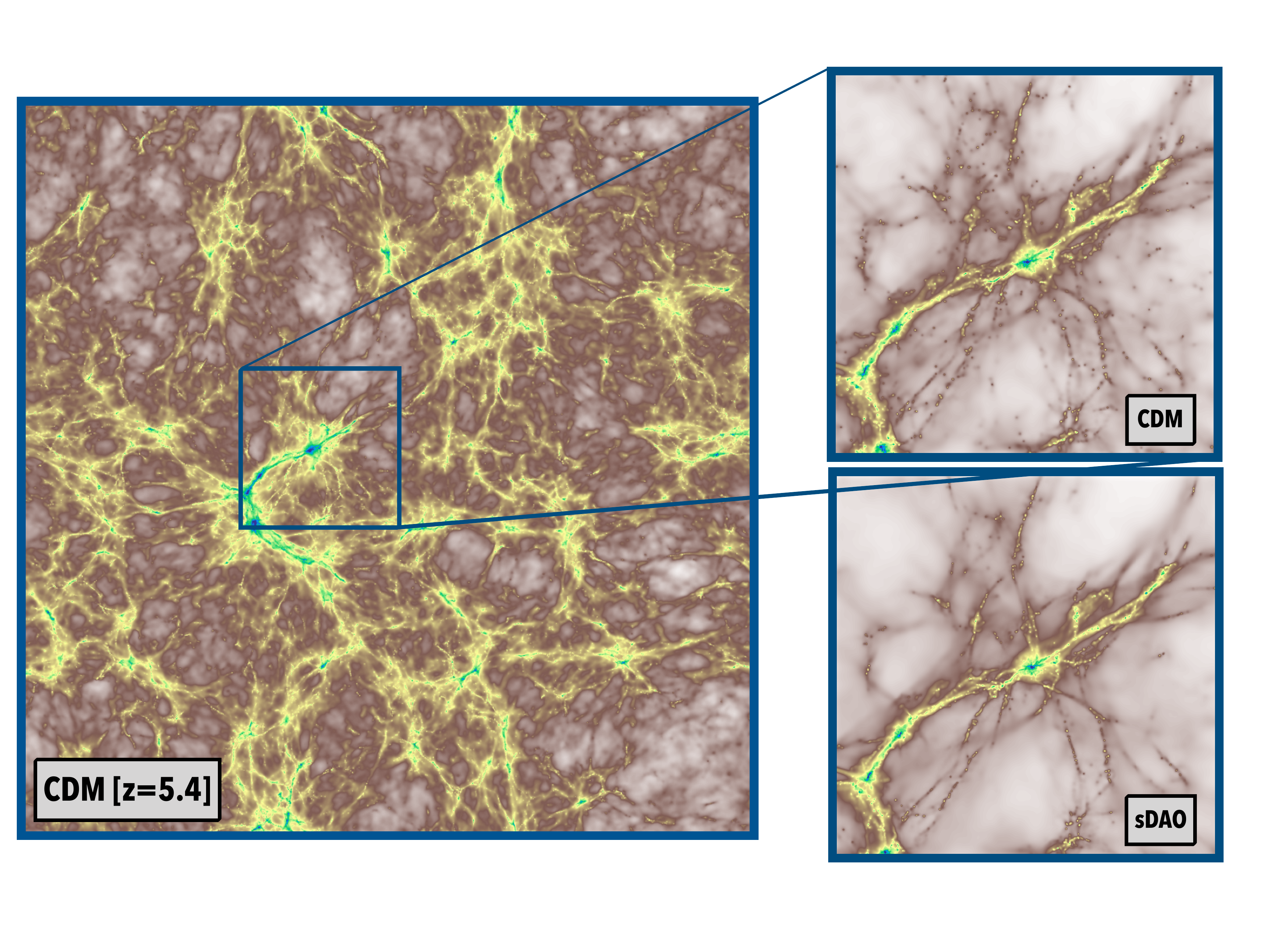}
    \caption{Images of the gas density at $z=5.4$ obtained from our hydrodynamical simulations. On the left, we project the simulation box along the $z$-axis in a projection of comoving dimensions $\left(20\times20\times4\right)\,h^{-1}$cMpc. The smaller panels zoom into a region centred on the most massive halo at this redshift in a window of size $\left(4\times4\times2\right)\,h^{-1}$cMpc in CDM (upper right) and the equivalent region in \ethostwo{} (lower right). While differences are hard to discern on these scales, the small-scale cutoff in the \ethostwo{} model results in a smoother matter distribution than the CDM volume at the same epoch. These images were processed using the publicly-available {\sc Py-SPHViewer} package \citep{pysph}.}
    \label{fig:image_scales}
\end{figure*}
\begin{equation}
    G_t(r,\sigma) = \mathcal{N} \begin{cases}
        \exp{(-A^2 r^2)}\,, & {\rm for}\, r \leq h_{sml}\\
        0\,, & {\rm otherwise}
        \end{cases}
\label{eq:trunc}
\end{equation}
where: 
\begin{equation}
    \begin{split}
        \sigma^2 &= \frac{h_{sml}^2}{8\upi^{1/3}}\,,\\
        A^2 &= \frac{4\upi^{1/3}}{h_{sml}^2}\,,\\
        \mathcal{N} &= \frac{8}{\upi h_{sml}^3} \left[ {\rm erf} (t) - \frac{2t \exp{(-t^2)}}{\sqrt{\upi}}\right]^{-1}\,,\\
        t &= 2\upi^{1/6}\,,
    \end{split}
\end{equation}
and $r=h_{sml}$ is the radius at which the Gaussian kernel is truncated. In smoothed particle hydrodynamics (SPH) simulations, $h_{sml}$ is taken to be the gas particle's smoothing length, calculated using a fixed number of nearest-neighbours. In \arepo{}, gas is discretised in the form of Voronoi cells rather than SPH particles; we therefore define an `effective' smoothing length, $h_{sml,i}$, for each gas cell, $i$, as:
\begin{equation}
    h_{sml,i} = f \left( \frac{3 m_i}{4\upi\rho_i} \right)^{1/3}\,,
\end{equation}
where $m_i$ and $\rho_i$, respectively, are the mass and mass density of the gas cell, and $f=4$ is some normalisation factor. Our results are insensitive to the precise choice of $f$.

By dividing each line-of-sight into $N$ bins, we can compute the number density, $n_{{\rm H}}(j)$, \hi{}-weighted temperature, $T(j)$ and \hi{}-weighted peculiar velocity field, $v(j)$, at each bin $j$ (in velocity space) using only the subset of gas cells that intersect each ray. Following exactly the methodology laid out in \cite{Theuns1998}, we can then calculate the optical depth, $\tau(k)$, for the $k$th velocity bin along the line-of-sight as:
\begin{equation}
\begin{split}
\tau (k) &=\sum_j \sigma_\alpha \frac{c}{V_{{\rm H}}(j)} n_{{\rm H}}(j) \, \Delta \times \frac{1}{\sqrt{\upi}} \exp{ \left( -\left[\frac{v(k)-v(j)}{V_{{\rm H}}(j)} \right]^2 \right) }\,, 
\\ V_{{\rm H}}^2(j) &= 2k_{{\rm B}} T(j)/m_{{\rm H}}\,,
\end{split}
\end{equation}
where $c$ is the speed of light, $\Delta$ is the width of each bin in units of physical distance, $x$, along the line-of-sight, $k_{{\rm B}}$ is the Boltzmann constant and $\sigma_\alpha = 4.45 \times 10^{-18}$ cm$^2$ is the cross-section of the hydrogen Lyman-$\alpha$ transition. The corresponding transmitted flux is then given by $F=e^{-\tau}$, where $\tau$ is the integrated optical depth along this line-of-sight.

Due to the considerable uncertainty about the level of photoionisation, we follow the standard procedure of rescaling our simulated spectra to the observed optical depth at the corresponding redshifts. At low redshift, $z\sim2-3$, the rescaling factor is small; at high redshift, $z\sim4-6$, however, the rescaling becomes increasingly important as fluctuations in the assumed UV background start to become an issue. In particular, at each simulation output, we rescale the optical depths of simulated spectra such that the {\it mean} transmitted flux matches the {\it observed} mean flux at that redshift. For the observed mean fluxes, we use the values reported by \cite{Walther2018} for $z<4$ and by \cite{Viel2013} for $z\geq 4$. The factor by which the CDM and \ethostwo{} spectra are rescaled are not too dissimilar at $z>3$, and are almost identical at lower redshift (see Fig.~\ref{fig:meanfluxz}).

This rescaling procedure is widespread in the simulation community and its validity is worth reflecting on for a moment. As we have mentioned previously, the motivation for rescaling the optical depth is the uncertainty of the photoionisation rate. In practice, one assumes that the \hi{} abundance is directly proportional to the photoionisation rate and hence, the optical depth can be rescaled by the same factor. The assumption implicitly neglects the following effects, which we consider in turn:
\begin{enumerate}
    \item Deviations from equilibrium: these only play an important role during reionisation and determine how strongly the gas is being heated. At the redshift of the observations of the Lyman-$\alpha$ forest, the ionisation degree should be back in equilibrium. The only remaining effect is the slightly enhanced temperature. 
    \item Collisional ionisation: this is a negligible effect at IGM temperatures.
    \item A constant recombination rate: since the \hi{} fraction is of order $\sim10^{-4}$, the \hii{} fraction and hence the recombination rate practically do not change for small variations of the \hi{} abundance.
    \item Spatial fluctuations of the photoionisation rate: these disappear quickly after the end of reionisation. For very high gas densities, e.g. in damped Lyman-$\alpha$ systems (DLAs), one would have to take into account self-shielding of gas. 
\end{enumerate}
In summary, the rescaling procedure is valid for the optical depths and temperatures of the IGM \citep[see, e.g., the discussions in][]{Rauch1997,Weinberg1997,Theuns1998,Bolton2005}. At $z>5.5$, the assumption of a homogeneous photoionisation rate fails severely and it is then necessary to perform radiative transfer calculations. The validity of rescaling the mean flux at these high redshifts then becomes questionable. In what follows, we limit our analysis to $z\leq5.4$.

\begin{figure}
    \centering
    \includegraphics[width=\columnwidth]{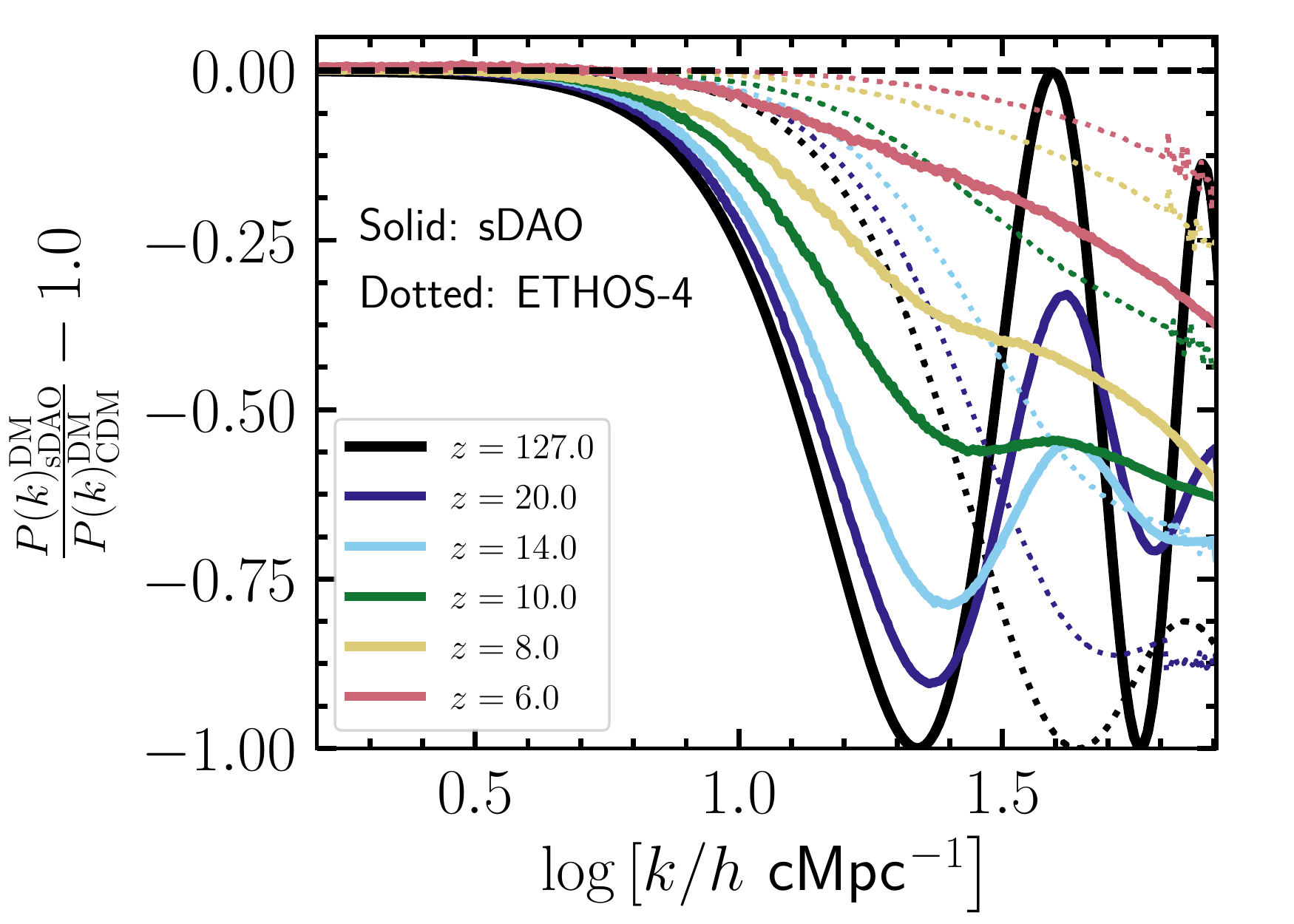}
    \caption{Redshift evolution of the ratio of the non-linear matter power spectrum at $z=20,14,10,8,6$ for the simulations presented in this paper. The power spectra are measured using only the distribution of DM particles in the simulation snapshots. For comparison, we also show the ratio of the {\it linear} power spectra used as the input to making initial conditions at $z=127$; we also show results for the {\sc Ethos-4} model with dotted lines. While the matter distribution shows significant differences between the \ethostwo{} and CDM model at high redshift (including the signature of DAOs), these differences are suppressed at lower redshift. In particular, the DAOs are no longer visible in the DM distribution at $z=6$. On the other hand, DAOs are nearly smoothed as early as $z=20$ in the {\sc Ethos-4} model due to our finite numerical resolution.}
    \label{fig:nlPk}
\end{figure}

\begin{figure*}
\centering
\includegraphics[width=0.48\textwidth]{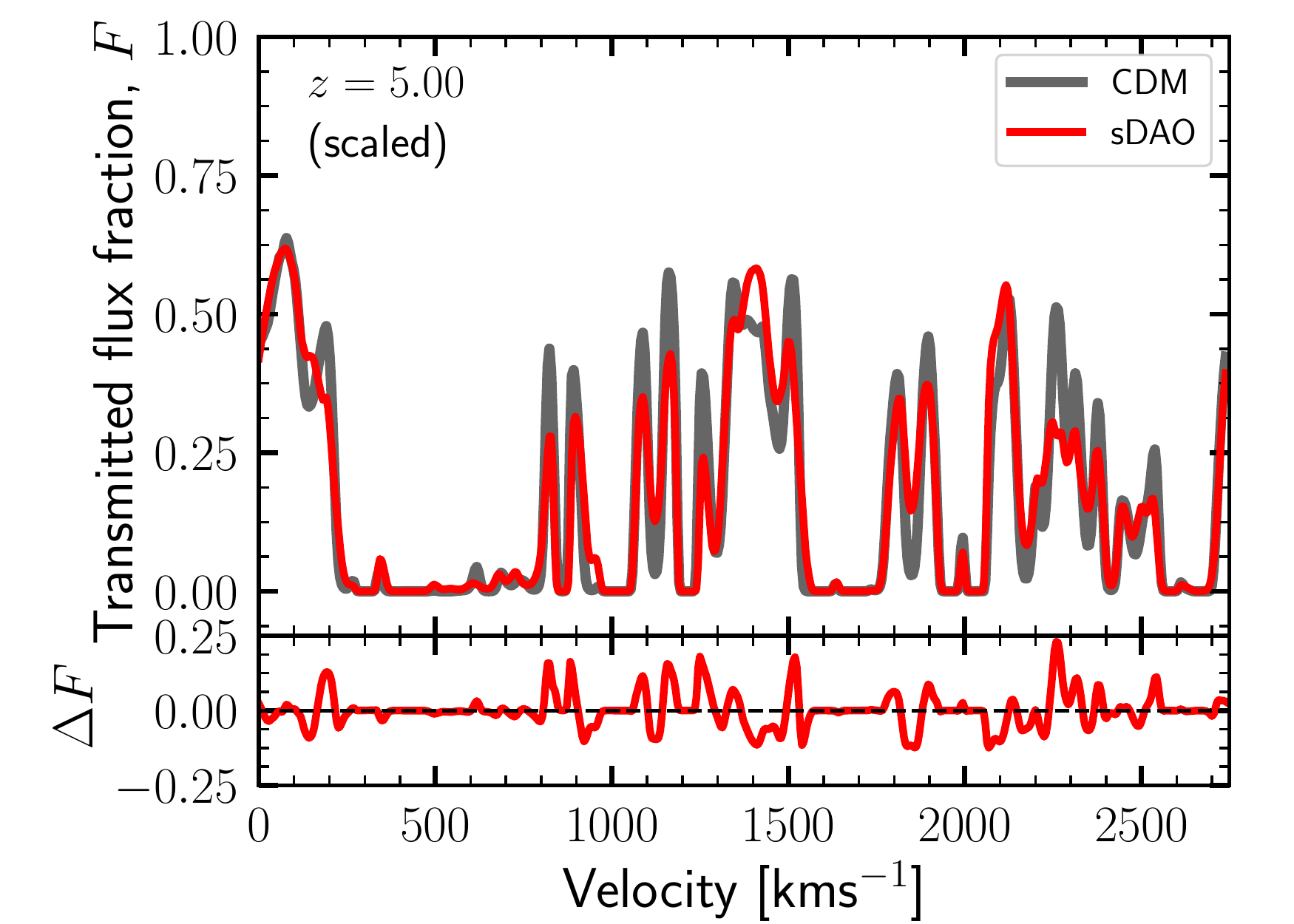}
\includegraphics[width=0.48\textwidth]{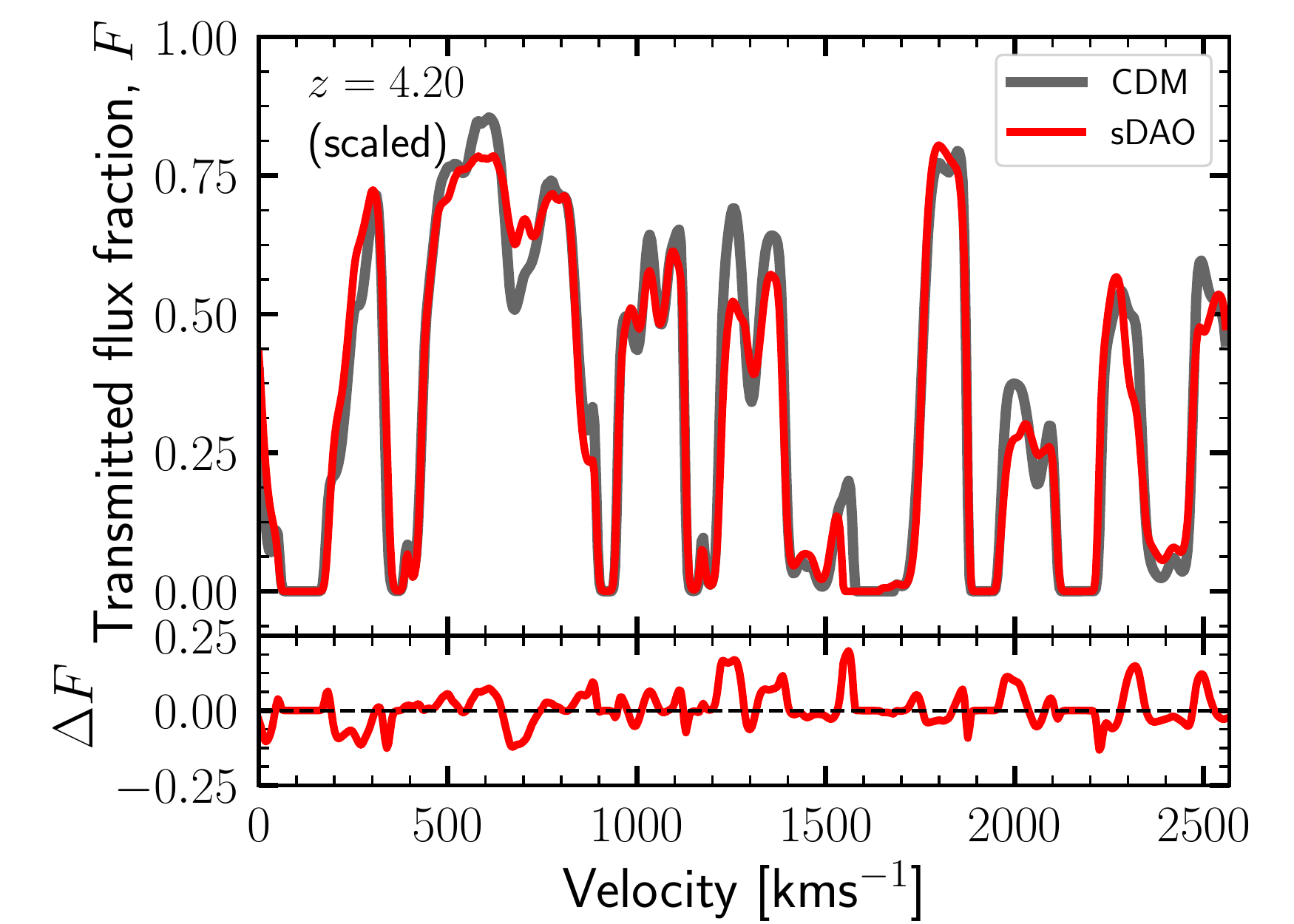}\\
\includegraphics[width=0.48\textwidth]{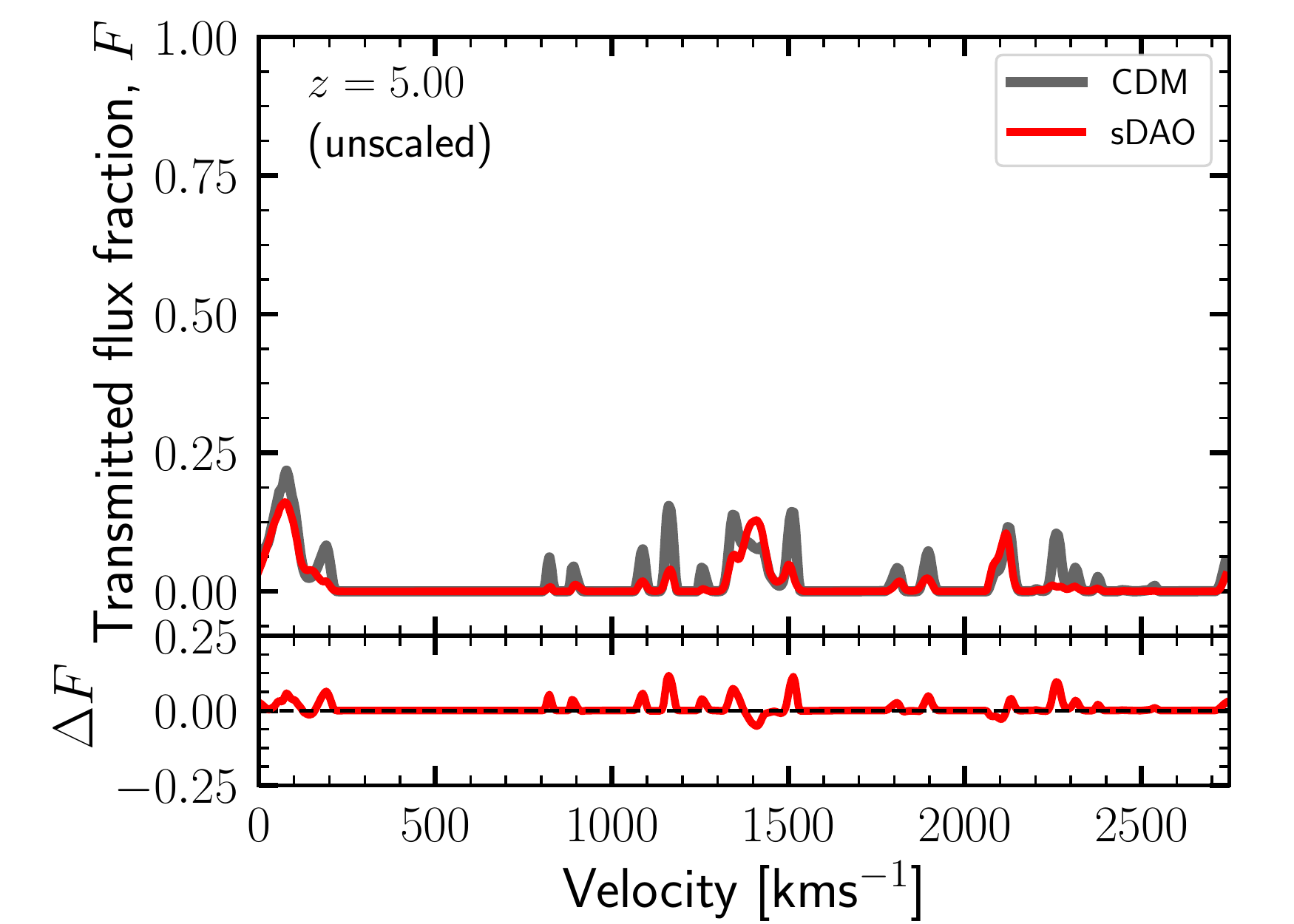}
\includegraphics[width=0.48\textwidth]{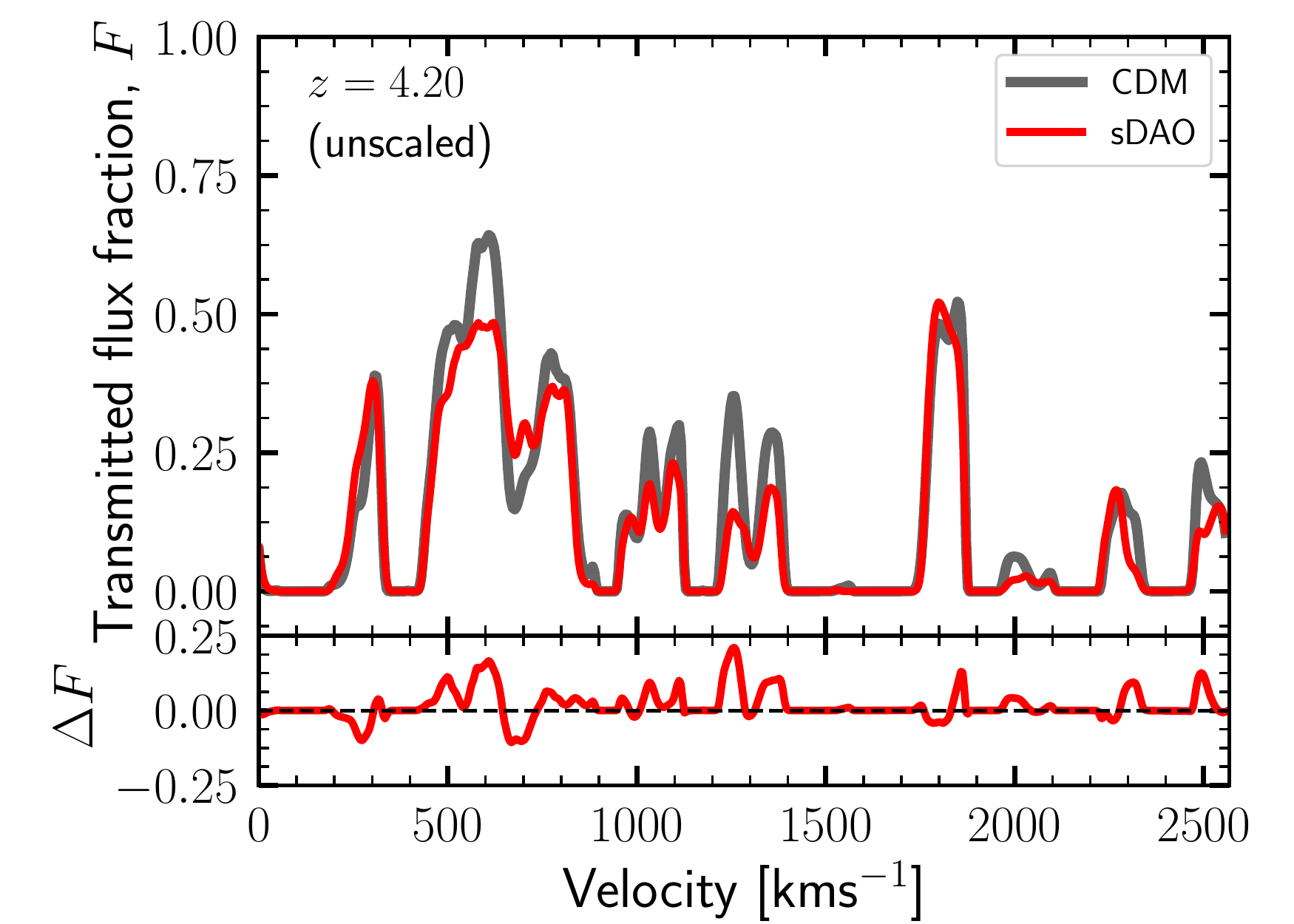}\\
\includegraphics[width=0.48\textwidth]{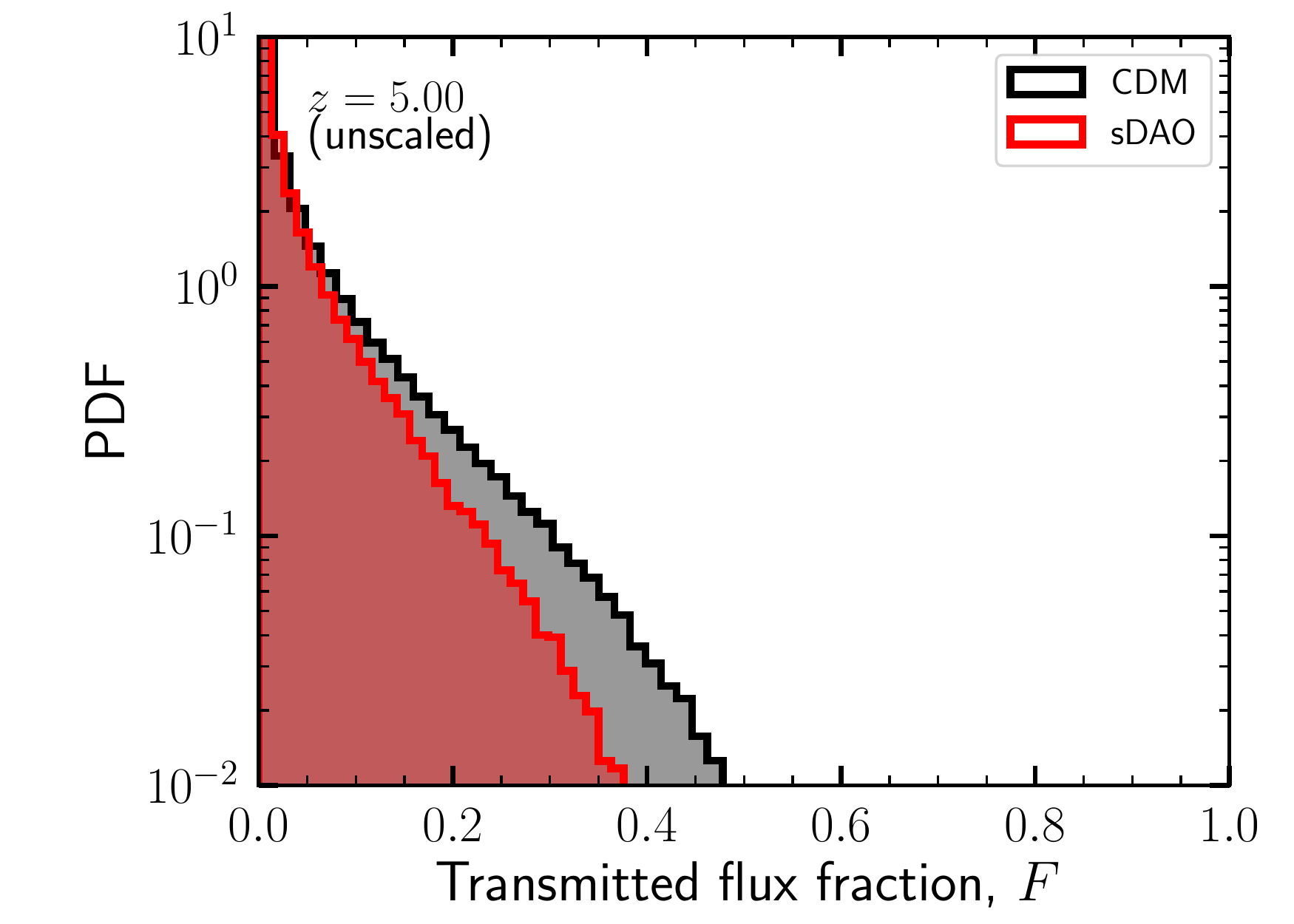}
\includegraphics[width=0.48\textwidth]{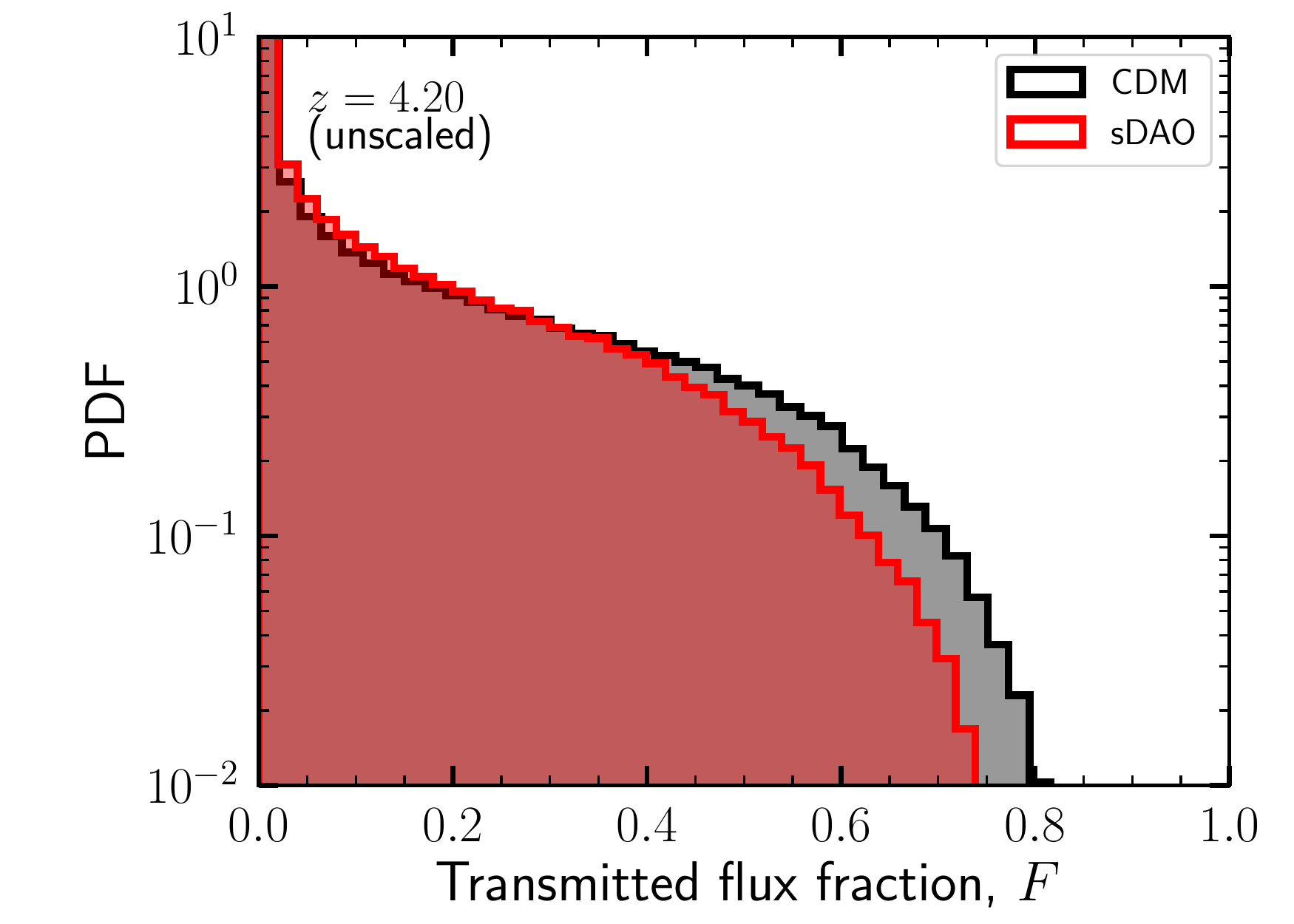}
\caption{{\it Top row}: Synthetic mock spectra extracted from our simulations at $z=5$ (left) and $z=4.2$ (right). Each panel shows the transmitted flux fraction, $F$, in velocity space for a specific randomly-selected line-of-sight through the simulation volume at the corresponding redshift. The lines-of-sight are chosen such that the same spatial region is probed in the CDM (black) and \ethostwo{} (red) simulation volumes. Clearly, more intervening structure can be inferred from the CDM spectra as evidenced by the deeper transmission lines than in \ethostwo{}. The lower panels show the relative difference in transmitted flux i.e., $\Delta F = F_{{\rm CDM}} - F_{{\rm sDAO}}$. {\it Middle row}: The same lines-of-sight shown in the top row but before we rescale the mean flux to the observed values. {\it Bottom row}: The unscaled flux PDF at these redshifts. These panels show the ``true'' difference in the transmitted flux between the CDM and \ethostwo{} models i.e. by removing any artefacts that may be brought in by the different amount of rescaling required for the two DM models.}
\label{fig:los_spectrum}
\end{figure*}

\section{Results}
\label{sect:results}

\subsection{The clustering of matter}

As a precursor to the main analysis in this paper, we show in Fig.~\ref{fig:image_scales} the projected gas density map from our simulation volumes at $z=5.4$. The largest panel shows a $\left(20\times20\times4\right)\,h^{-1}$cMpc projection from the CDM simulation; the smaller panels zoom into a $\left(4\times4\times2\right)\,h^{-1}$cMpc region centred on the most massive halo at this time as it appears in the CDM (upper right) and \ethostwo{} (lower right) simulations. While general large-scale filaments and knots look identical in the two density maps, there is noticeable absence of small-scale structure in the \ethostwo{} image, in which the gas density distribution is smoother than in CDM. This situation is identical to what is observed in standard WDM simulations, in which the matter distribution is smoothed through free-streaming induced by the cutoff in the linear power spectrum, although the mechanism in operation here is collisional Silk damping, rather than free-streaming. The smoothed gas distribution in WDM models is manifest as a cutoff in the Lyman-$\alpha$ flux spectrum at small-scales; our aim in the subsequent sections is to investigate if the {\it resurgence} in power at small-scales -- predicted by models with strong DAOs, but not by thermal relic WDM -- can be probed by the Lyman-$\alpha$ forest. 

Before examining the Lyman-$\alpha$ forest, it is instructive to first look at the DM distribution predicted by these models. Fig.~\ref{fig:nlPk} shows ratios (\ethostwo{} to CDM) of the non-linear DM power spectrum at $z=20,14,10,8$ and $6$ (coloured lines), measured directly from the DM particles in each simulation at the corresponding redshifts. For comparison, we also show, in black, the ratio of the {\it linear} power spectra in these models, which were used to generate the initial conditions at $z=127$. The {\sc Ethos-4} model, comparable to a 3.3 keV thermal relic, is represented by the dotted lines.

From Fig.~\ref{fig:nlPk}, one clearly notices that the characteristic DAO peaks are very prominent at early times. At $z=20$ and $z=14$, the first DAO peak is still noticeable in the DM distribution (at $\log [ k /h$cMpc$^{-1}] \approx 1.6$), but only marginally so by $z=10$. As gravitational collapse continues at $z<10$, increasing the overall power on all scales, the DAO peak is gradually washed away as a result of mode coupling in the (weakly) non-linear regime of structure formation. By $z=6$, any signature of DAOs has completely disappeared -- qualitatively, the ratio of the non-linear power spectra looks more similar to an ordinary thermal relic WDM particle. This is consistent with the findings of \cite{Buckley2014} and \cite{Vogelsberger2016} who also noted the absence of acoustic peaks in the DM distribution at relatively high redshift ($z\sim6-8$). This is particularly true for the {\sc Ethos-4} model, where the DAOs are nearly absent as early as $z=20$ (given the finite resolution of our numerical setup).

\begin{figure}
    \centering
    \includegraphics[width=\columnwidth]{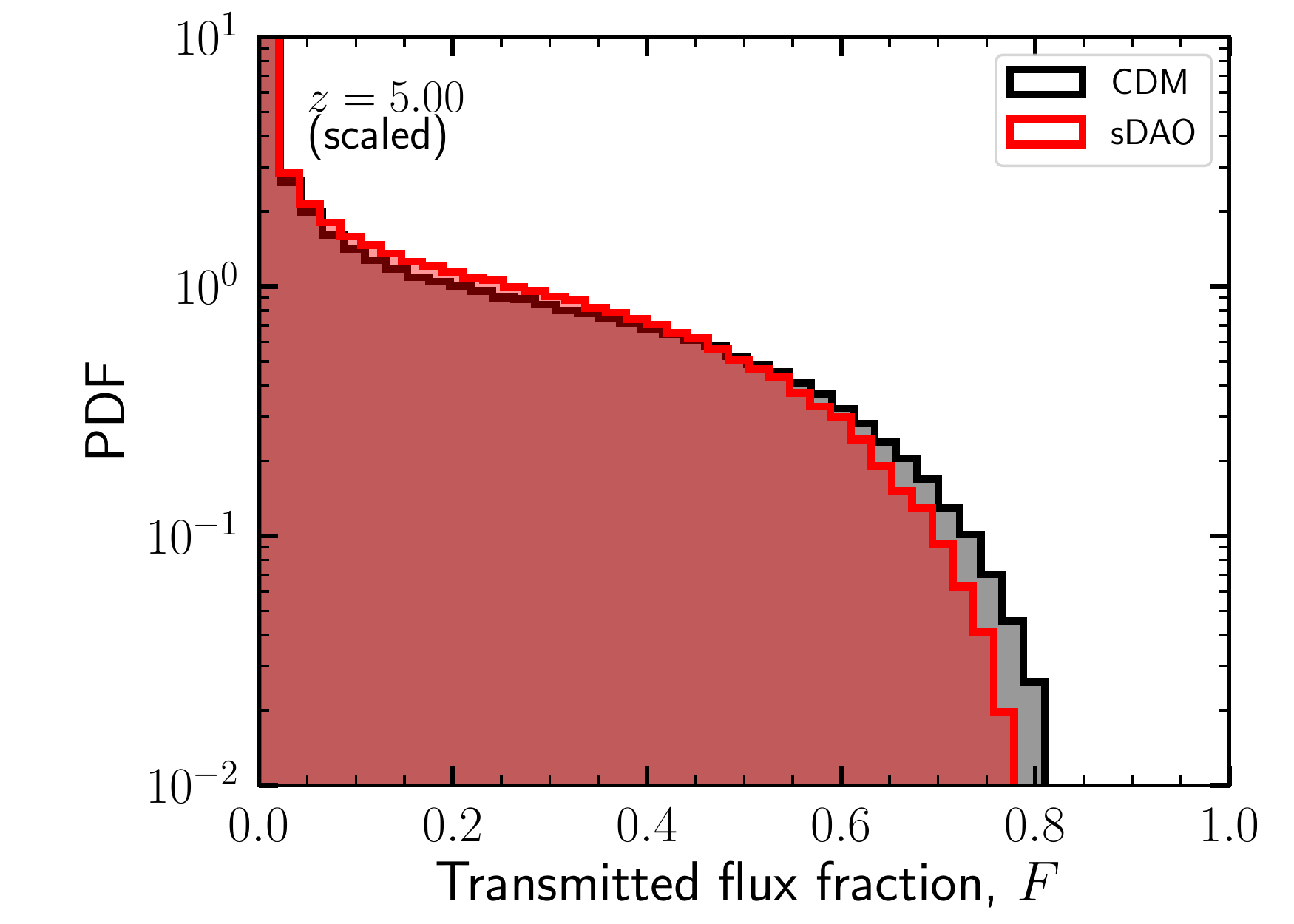}\\
    \includegraphics[width=\columnwidth]{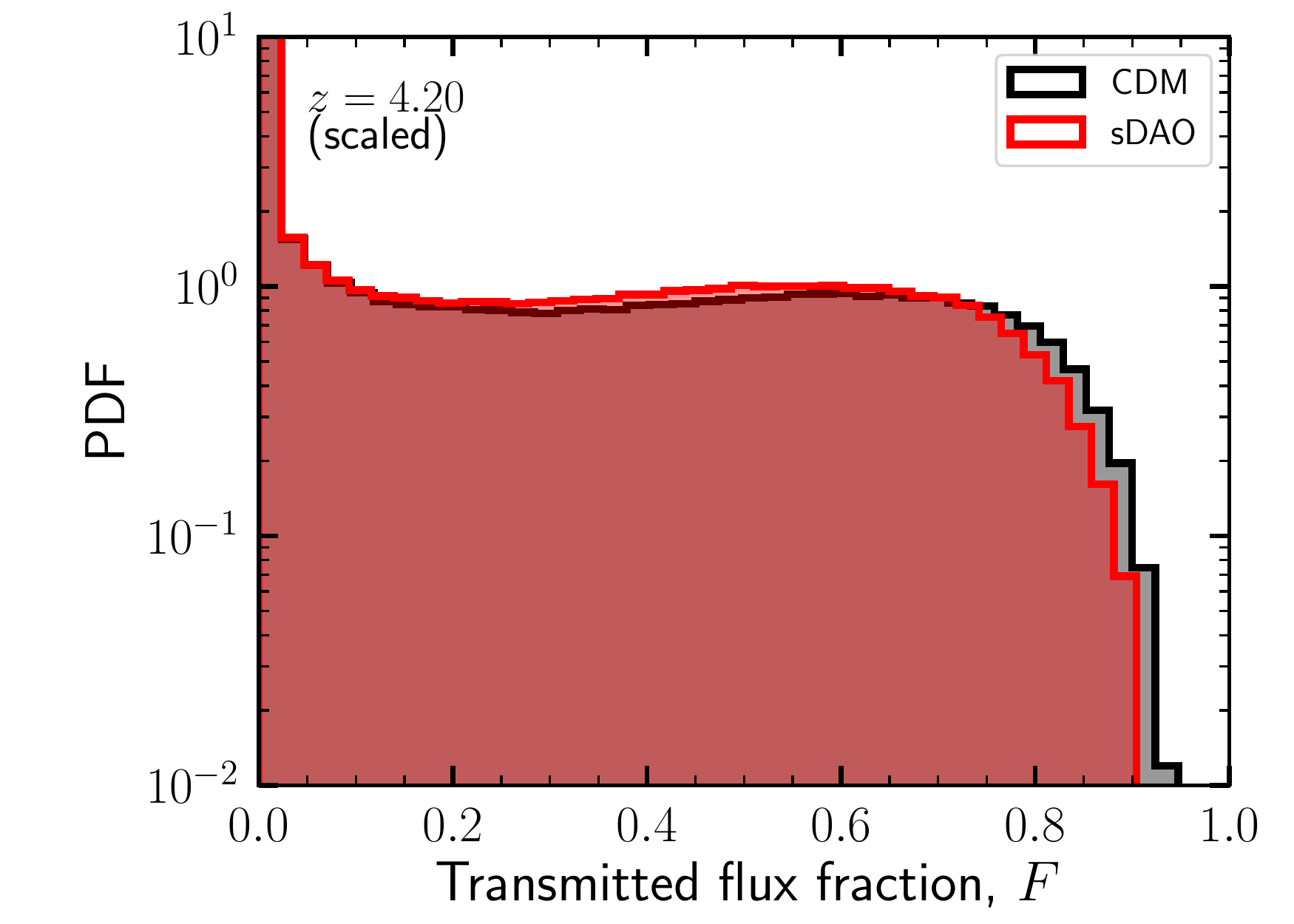}\\
    \includegraphics[width=\columnwidth]{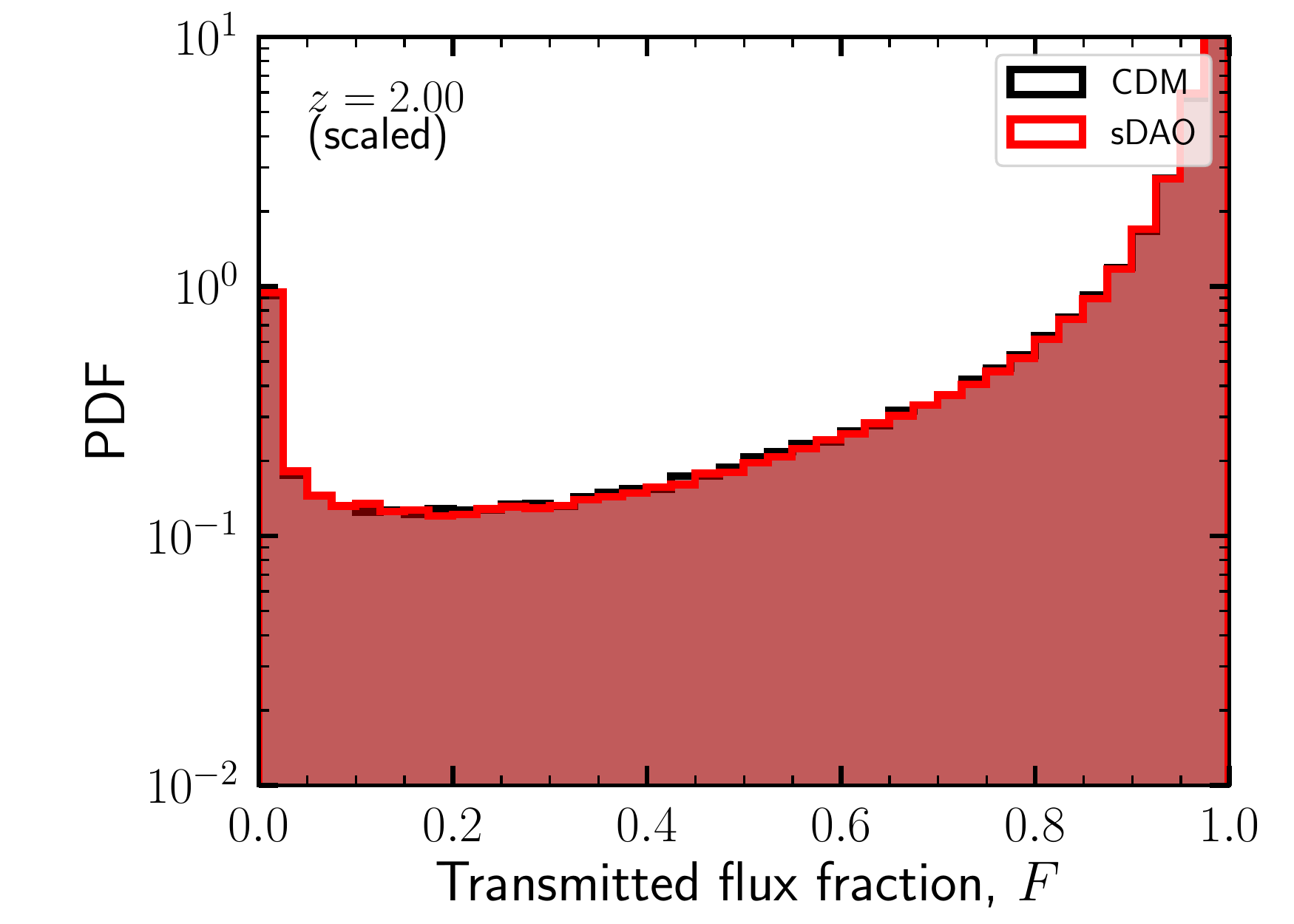}
    \caption{Comparison of the PDF of the transmitted flux, $F$, at $z=5,4.2,2$ (top to bottom) after rescaling the mean transmitted flux. While both CDM and \ethostwo{} show identical PDFs at low redshift, the CDM simulations display an extended tail of high flux at higher redshift.}
    \label{fig:fluxPDF}
\end{figure}

\subsection{The Lyman-$\alpha$ forest at a glance}

In the top row of Fig.~\ref{fig:los_spectrum} we present the transmitted flux fraction, $F$, measured along random lines-of-sight at $z=5$ (left) and $z=4.2$ (right) for the CDM and \ethostwo{} models. The spectra have been created following the procedure outlined in Section~\ref{sect:lya}, after rescaling the mean transmission in each simulation box to the observed transmitted flux. In each case, the lines-of-sight have been chosen so as to probe the same spatial region in the two simulation volumes. The lower sub-panels show the difference in the transmitted flux, $\Delta F = F_{{\rm CDM}} - F_{{\rm sDAO}}$. This figure highlights the fact that the same line-of-sight probes different intervening structure in the IGM of the two simulations. In particular, it is clear that, just as in WDM, absorption lines are in general deeper for a random line-of-sight in the CDM simulation than for \ethostwo{} model, signifying the presence of a more clumpy IGM. This difference is a direct consequence of the cutoff in the initial power spectrum. 

The \ethostwo{} and CDM models have been rescaled by different amounts when matching the simulated spectra to the observed mean transmitted flux. To disentangle the implications of rescaling from the different line shapes due to the modified cosmology at a given UV background, the middle row of Fig.~\ref{fig:los_spectrum} shows the {\it unscaled} line-of-sight spectra at $z=5$ and $z=4.2$. Clearly, there are residual differences between the two models even in the unscaled case. Quantitatively, these effects are seen more clearly in the bottom row of Fig.~\ref{fig:los_spectrum}, which shows the PDF of the unscaled transmitted flux for the two models at these redshifts. Considering just CDM to begin with, it is noticeable that as time proceeds, there is a cutoff in the flux PDF at higher values of $F$, which comes about due to a combination of two effects: i) as the universe expands, the background density drops and a given overdensity needs to be larger (in linear dimensions) in order to produce the same signal strength in the Lyman-$\alpha$ forest, and ii) as structure formation proceeds, the non-linear length scale moves to larger scales, implying that perturbations with larger wavelengths start to collapse, yielding a more structured (clustered) universe. As a result, the extended tail to large $F$ builds as gravitational collapse proceeds through cosmic time. There is a clear extended tail of high flux in CDM which is less prominent in \ethostwo{}. This can be ascribed to the delayed collapse of the first haloes as a result of the suppressed small-scale density fluctuations induced by the DM-radiation coupling in the \ethostwo{} model.

A more realistic comparison of the two models is shown in Fig.~\ref{fig:fluxPDF}, which displays the {\it scaled} flux PDFs as a function of redshift. The panels, from top to bottom, show the flux PDFs at $z=5, 4.2$ and $2$. This figure highlights a qualitative difference in the manner in which structure formation proceeds in the \ethostwo{} model compared to CDM. At $z=5$, for example, the flux PDF is truncated at somewhat lower values of $F$ than it is in the CDM simulation. This can be ascribed to the delayed collapse of the first haloes as a result of the suppressed small-scale density fluctuations induced by the DM-radiation coupling in the \ethostwo{} model. The same difference, though smaller, is also manifest in the flux PDF at $z=4.2$. By $z=2$, however, the flux PDFs are almost indistinguishable between the two models. This is one of the generic features of models that exhibit a primordial cutoff in the linear power spectrum: while the formation of the first galaxies is delayed, structure formation proceeds more rapidly than in CDM afterwards. For the case of WDM, this has been demonstrated in detail in \cite{Bose2016b,Bose2017}. As we show in the following subsection, the same qualitative behaviour is manifest in the 1D flux power spectrum as well. 

\begin{figure}
    \centering
    \includegraphics[width=\columnwidth]{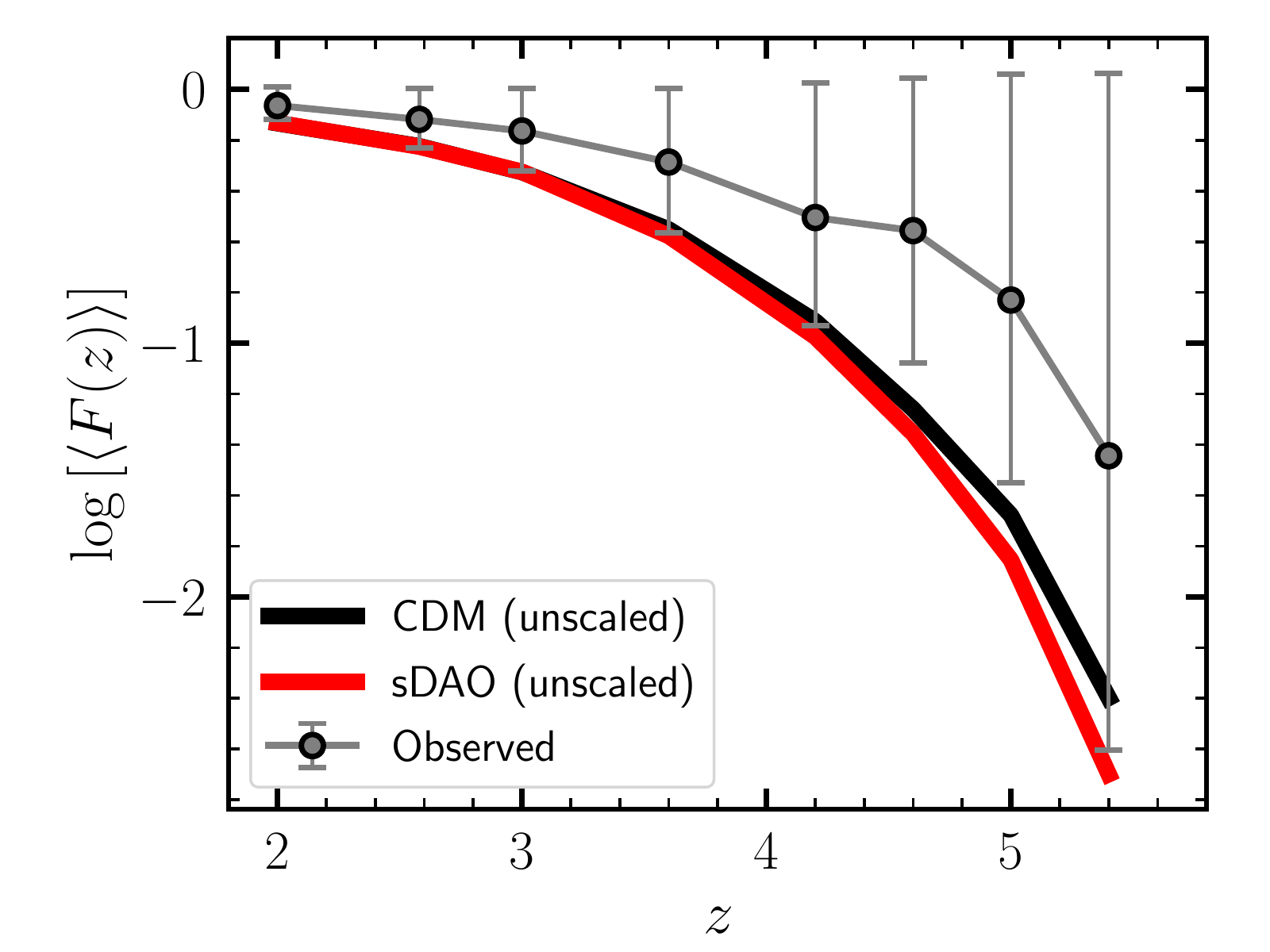}
    \caption{Redshift evolution of the (unscaled) mean transmitted flux in the two DM models compared to observed value of the mean transmission at each redshift. While \ethostwo{} spectra are rescaled by a larger factor than CDM spectra at $z\geq4$, the rescaling is almost identical towards lower redshift. Data obtained from \citet{Viel2013} and \citet{Walther2018}.}
    \label{fig:meanfluxz}
\end{figure}

Fig.~\ref{fig:meanfluxz} displays another way of visualising the different redshift evolution in the CDM and \ethostwo{} models. Here we show the {\it unscaled} mean transmitted flux as a function of redshift, compared to the observed values that we rescale to. Mean fluxes in the \ethostwo{} model are systematically lower at $z\geq4$ but nearly identical at later times. Figs.~\ref{fig:fluxPDF} and~\ref{fig:meanfluxz} therefore show explicitly the effects of delayed structure formation in the \ethostwo{} model, a direct consequence of the intrinsic cutoff in the linear power spectrum.

\subsection{The 1D flux spectrum}
\label{sect:fluxPS}

Next, we investigate if the distinctive feature of the \ethostwo{} model, the small-scale acoustic oscillation, is detectable in the Lyman-$\alpha$ forest. To probe this feature, we compute the 1D Lyman-$\alpha$ flux power spectrum. Following \cite{Viel2013}, at redshift $z$, we compute the power spectrum, $P_{{\rm 1D}} (k)$, of the {\it fractional transmission}, $\delta_F (z)$, which is defined as:
\begin{equation}
    \delta_F (z) = \frac{F(z) - \left< F(z) \right>}{\left< F(z) \right>}
\end{equation}
where $\left< F(z)\right>$ is the mean transmitted flux at redshift $z$. As described in Section~\ref{sect:lya}, the mean flux at every snapshot is rescaled to match the observed mean flux at that redshift. The power spectrum is calculated in this way for each of the 1024 lines-of-sight at a given redshift; the resulting value of $P_{{\rm 1D}} (k)$ at that redshift is then obtained by taking the mean value of the individual power spectra at each $k$ mode.

\begin{figure*}
    \centering
    \includegraphics[width=0.43\textwidth]{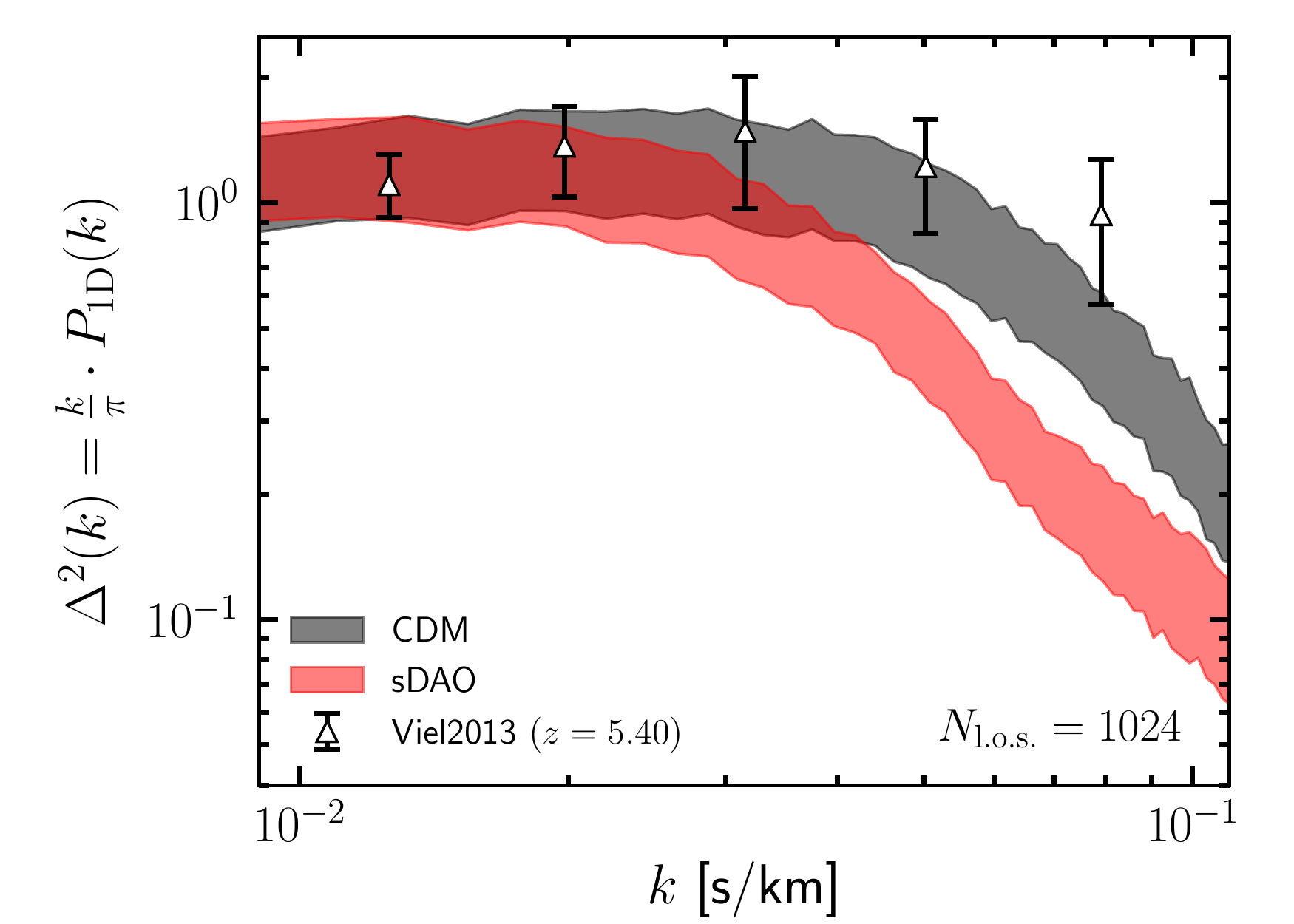}
    \includegraphics[width=0.43\textwidth]{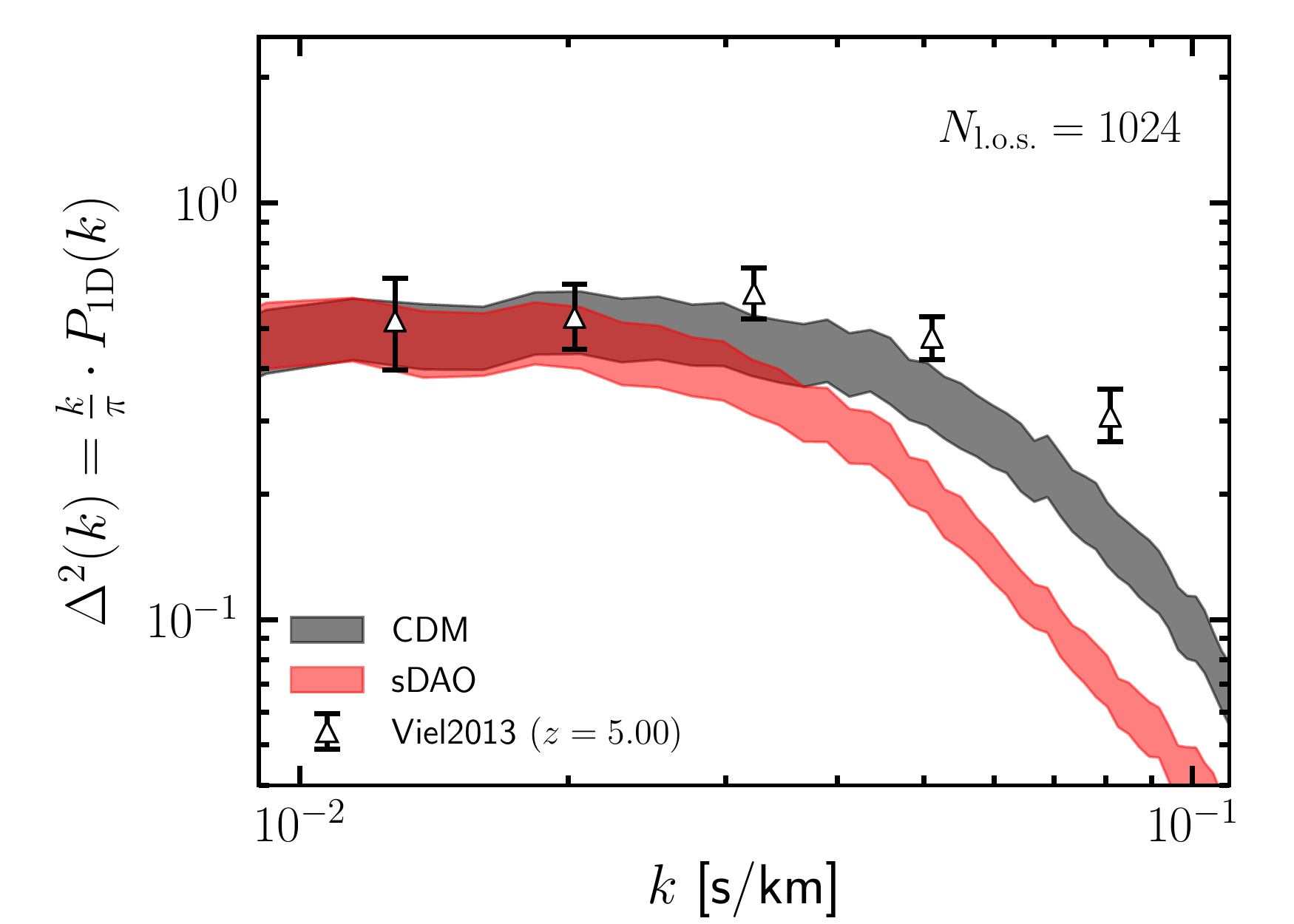}
    \includegraphics[width=0.43\textwidth]{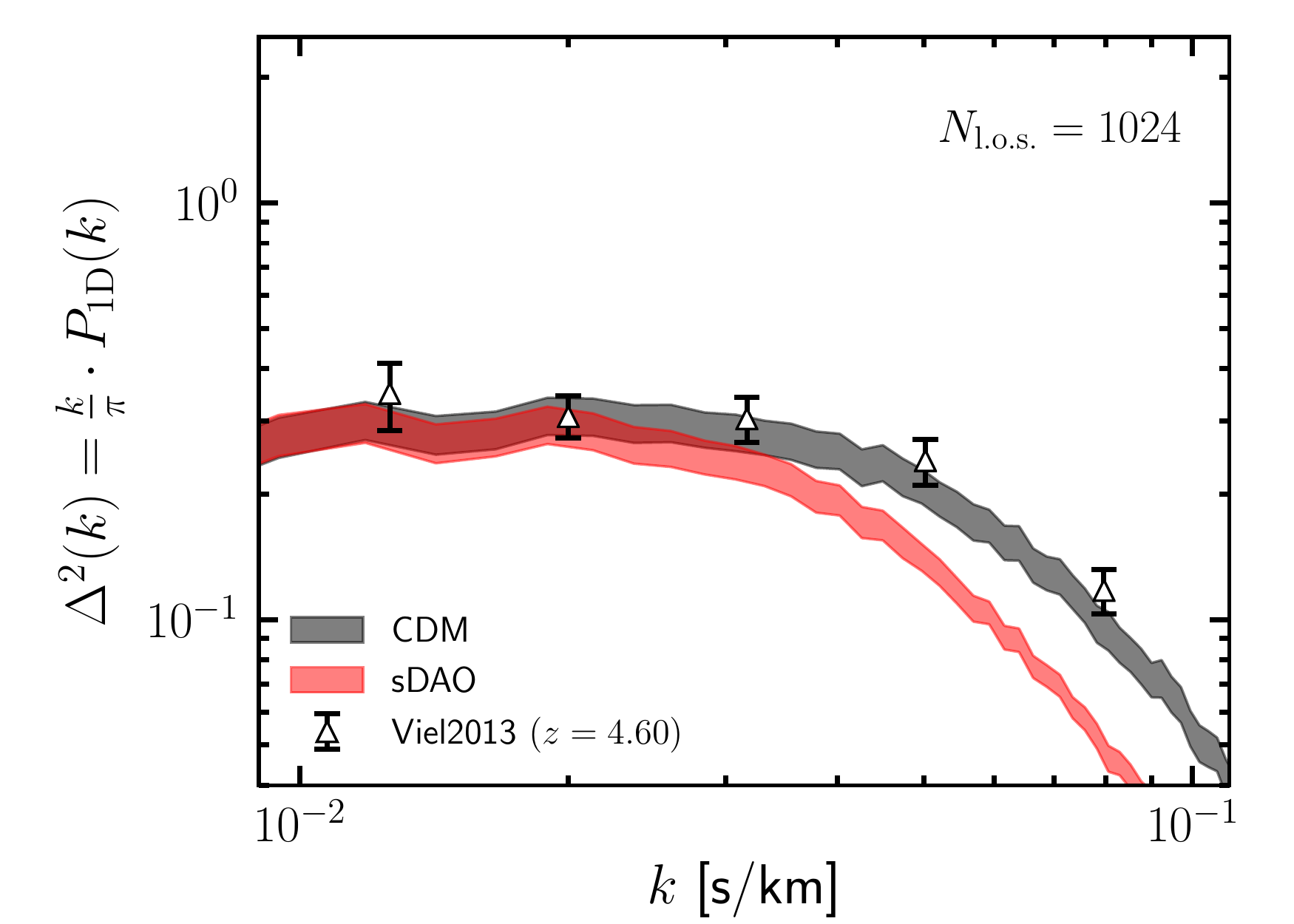}
    \includegraphics[width=0.43\textwidth]{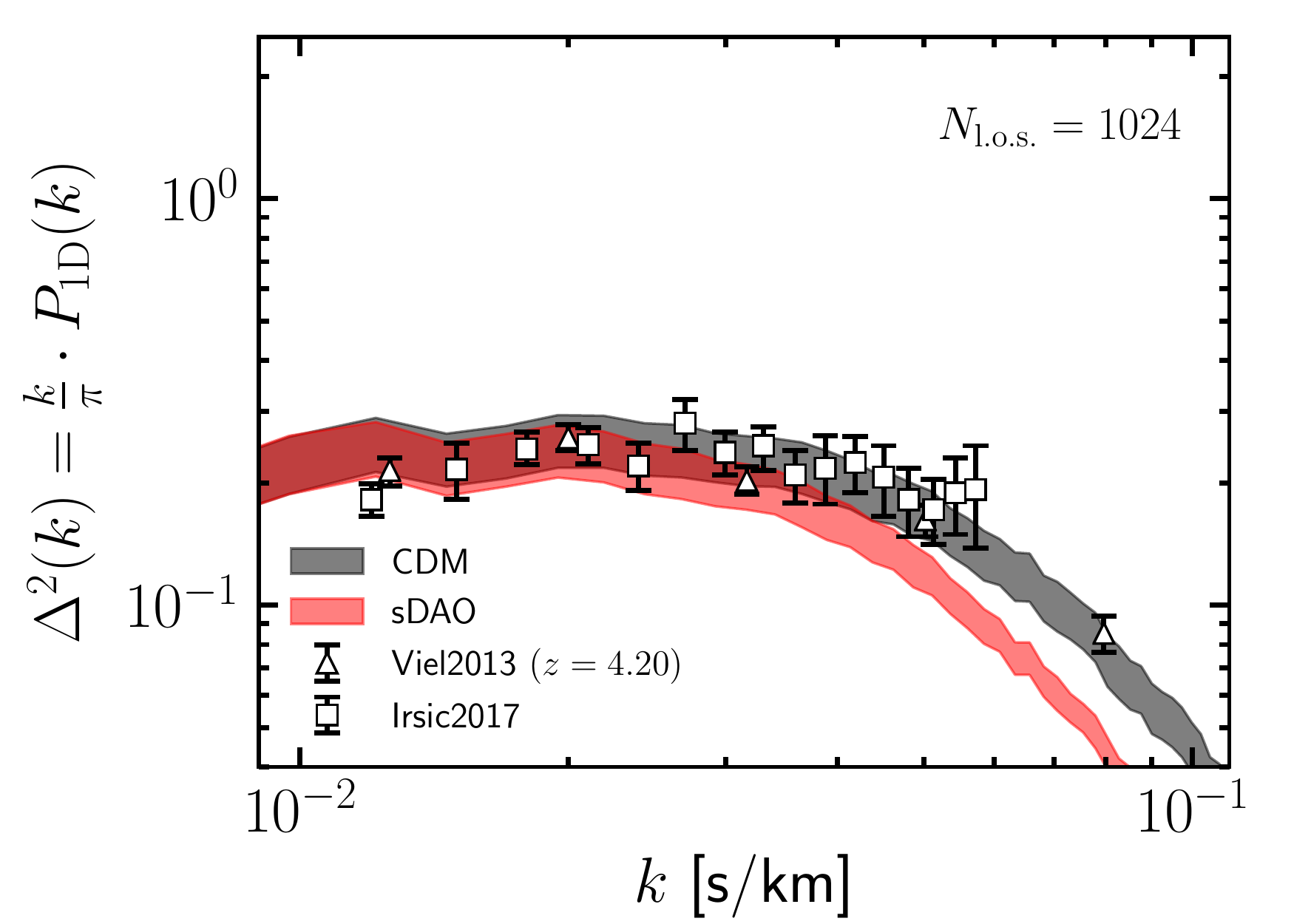} 
    \includegraphics[width=0.43\textwidth]{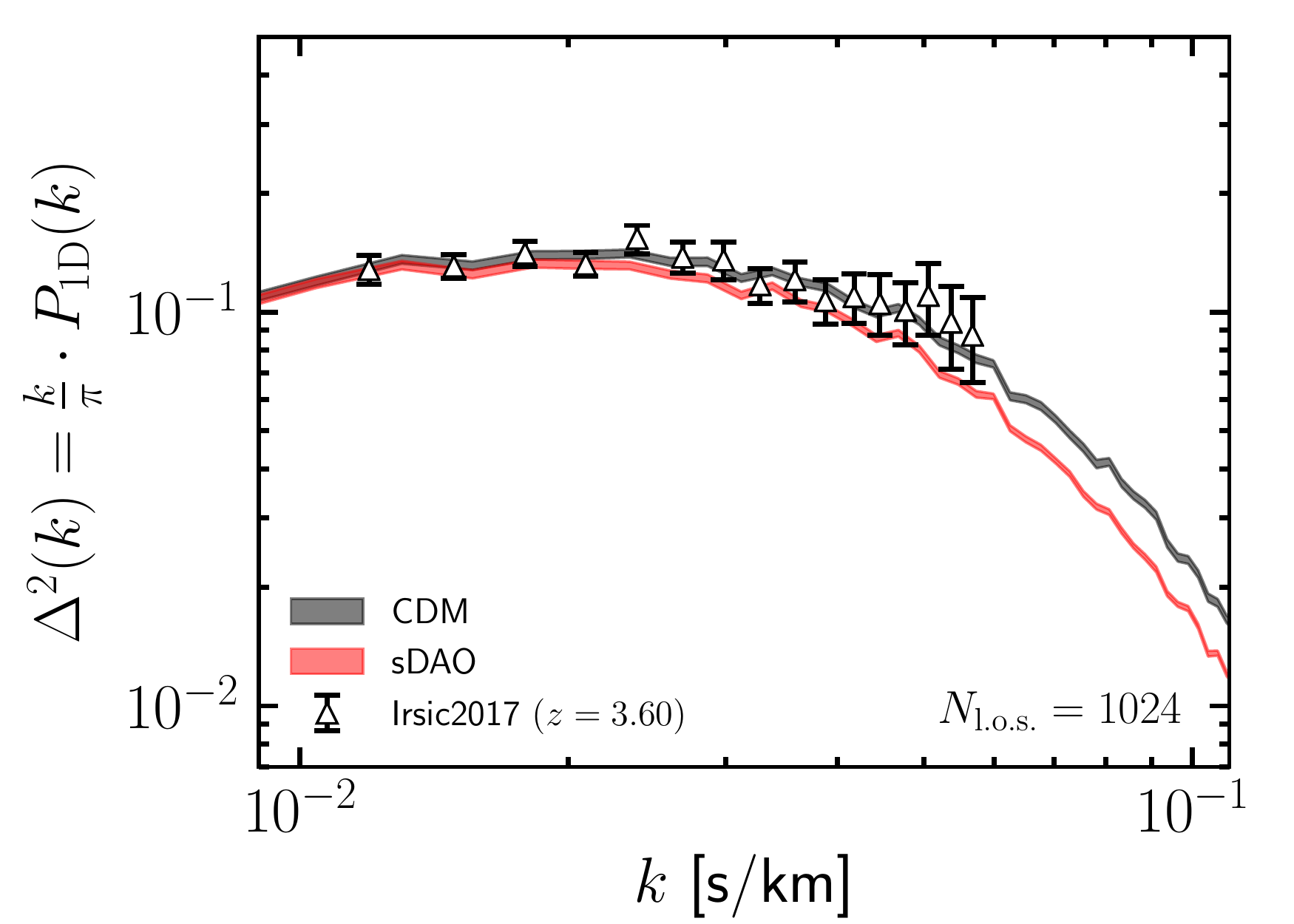}
    \includegraphics[width=0.43\textwidth]{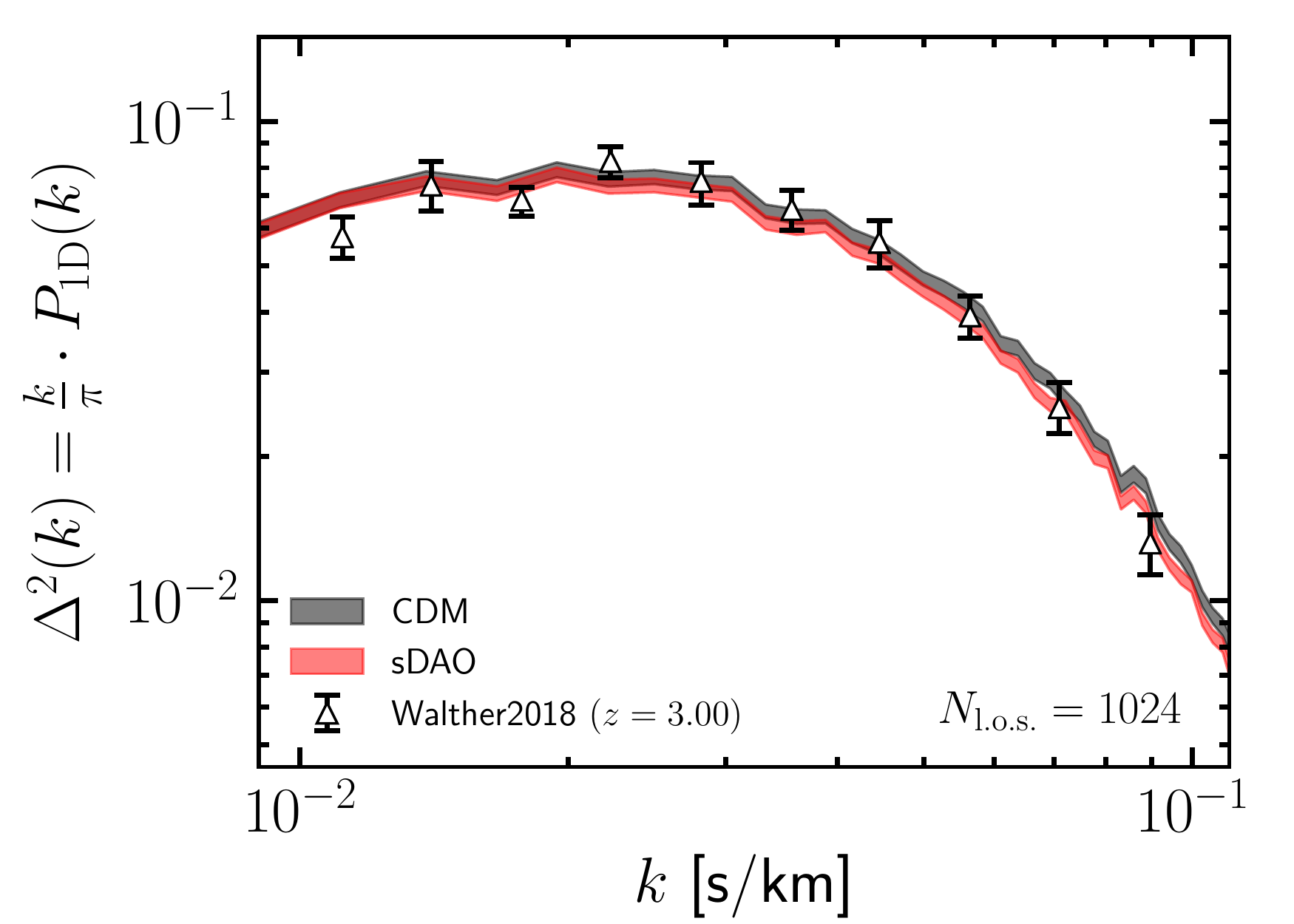}
    \includegraphics[width=0.43\textwidth]{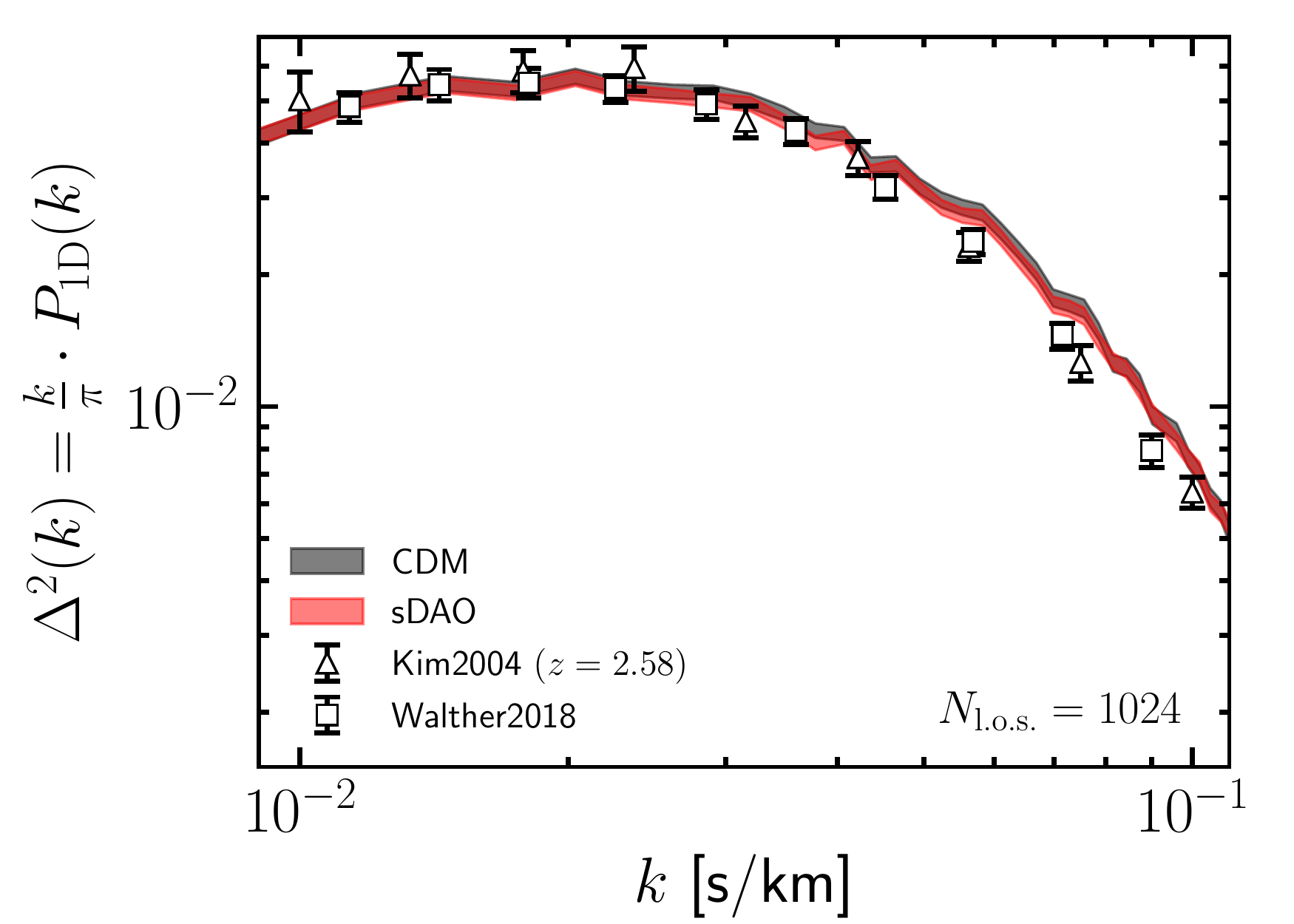}
    \includegraphics[width=0.43\textwidth]{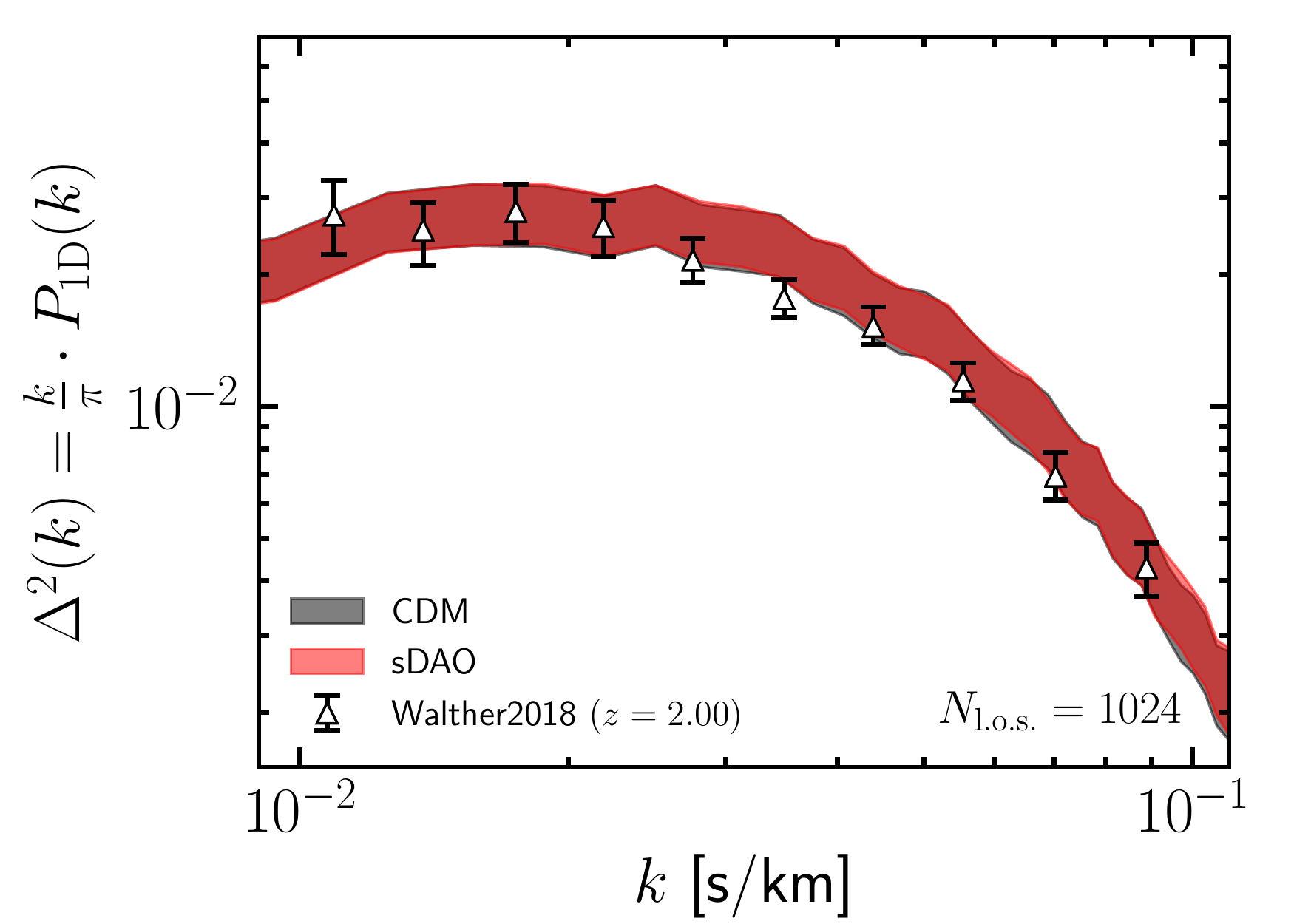}
    \caption{The 1D Lyman-$\alpha$ flux spectra for CDM and
      \ethostwo{}, compared to data obtained from MIKE/HIRES and XQ-100 quasar spectra at $z=5.4,5,4.6,4.2,3.6,3,2.58$ and $2$. To construct the simulated power spectra, we have, where possible, attempted to match the path length of the observed spectra. Each simulated spectrum makes use of 1024 lines-of-sight. The observational measurements are obtained from data compiled by \citet{Kim2004}, \citet{Viel2013}, \citet{Irsic2017b} and \citet{Walther2018}. The shaded regions encompass the reported uncertainty in observed mean transmission at that redshift, which translates to an uncertainty in the normalisation of the power spectra after rescaling.}
    \label{fig:obsCompare}
\end{figure*}

\begin{figure}
    \centering
    \includegraphics[width=\columnwidth]{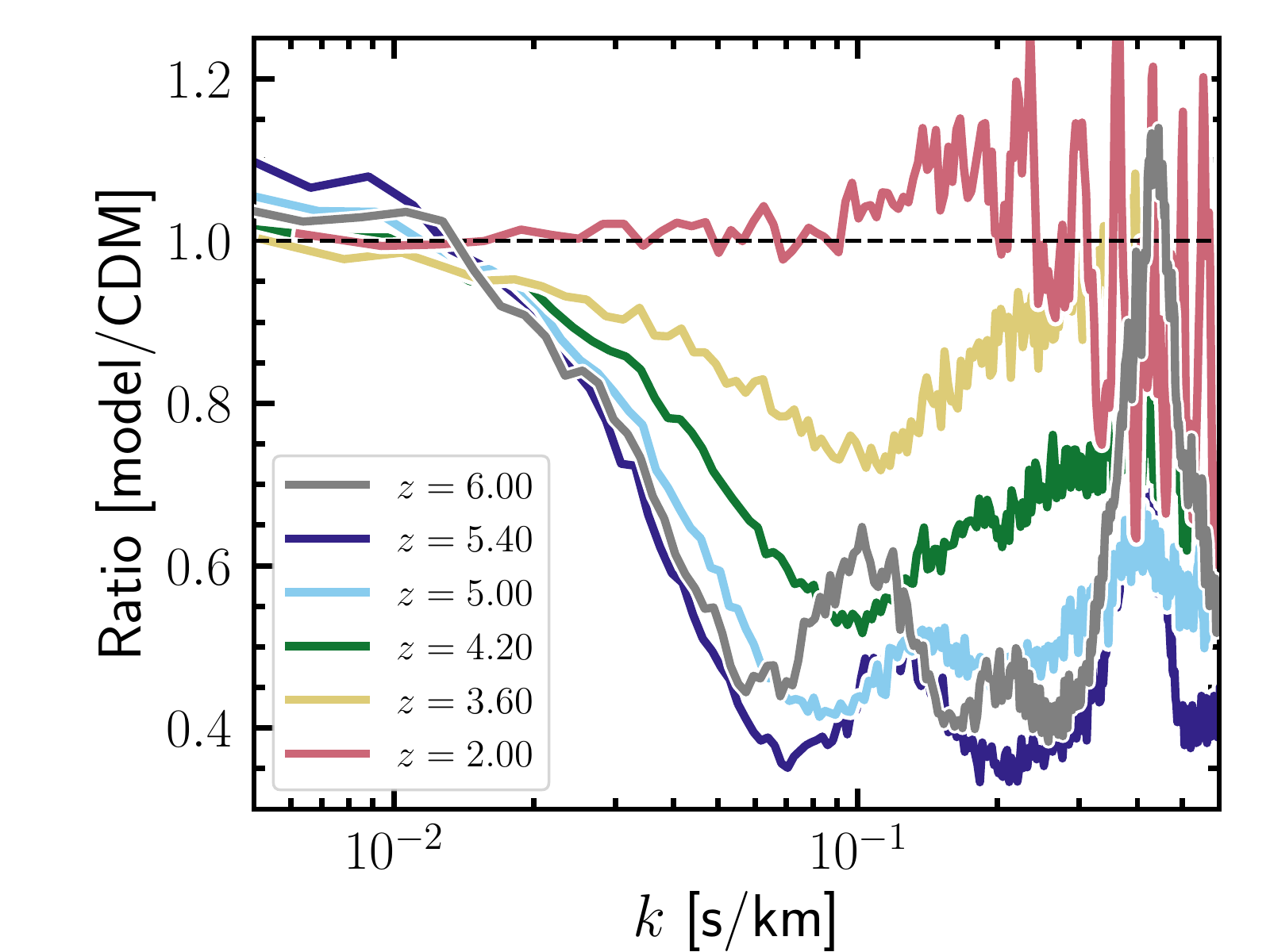}
        \caption{Ratio of the mean flux power spectra $\left[\Delta^2_{{\rm sDAO}} (k) / \Delta^2_{{\rm CDM}} (k) \right]$. For clarity, we do not show the observational data in this figure. The signature of DAOs (at $k=0.4$ s/km) can be noticed in the high redshift spectra, particularly at $z=6$ and $z=5.4$. Any evidence of DAOs is completely washed out by $z=2$. Note that the `bump' at $k\sim0.4$ s/km is numerical, and is set by the finite resolution of our simulation setup. This secondary feature is not sourced by DAOs.}
    \label{fig:ratioCompare}
\end{figure}

Fig.~\ref{fig:obsCompare} shows the results of this procedure for the CDM and \ethostwo{} simulations over a range of redshifts. Each of the models is represented by a shaded region which denotes the uncertainty in observed mean transmission at that redshift reported by \cite{Viel2013} ($z\geq4$) and \cite{Walther2018} ($z<4$), which we have propagated through to the normalisation of the simulated power spectra. In each panel we also display the observed flux spectra at each redshift, with data compiled by \cite{Kim2004}, \cite{Viel2013}, \cite{Irsic2017b} and \cite{Walther2018} from the MIKE/HIRES and XQ-100 quasar spectra samples. We do not include large-scale flux power spectrum measurements from BOSS \citep{Palanque2013} as the CDM and \ethostwo{} models are identical on these scales ($k \lesssim 0.02$ s/km).

The redshift evolution of the flux power spectra is reminiscent of the behaviour seen in Figs.~\ref{fig:nlPk} and~\ref{fig:fluxPDF}, in which the stark differences between the CDM and \ethostwo{} models diminish with redshift. For example, at $z=5.4$ the power spectra for the two models match only on the very largest scales ($k \leq 0.2$ s/km); in contrast, their power spectra are identical across all scales by $z=2$.

While the flux power spectra measurements at $z \leq 4.6$ obtained from the CDM simulation are in good agreement with the data even at the smallest scales (at least within the errors afforded by the uncertainty in the mean transmitted flux), this is not so for the two highest redshift bins ($z=5,5.4$). In particular, the simulated flux power spectra show a sharper decrement of power than is observed at scales smaller than $k\sim 0.07$ s/km. Part of this discrepancy may be due to incomplete masking of metal lines, which could add artificial power at small-scales. This effect would be more pronounced at higher redshift where masking all metal contributions is more challenging \citep[e.g.][]{Walther2018}. The bulk of the discrepancy, however, can be pinpointed to numerical resolution, as shown explicitly in the convergence tests performed by \citet[see their Fig. A4]{Bolton2017}. These authors show that at $z\sim5$, a simulation where each gas element is $\sim 10^6\,{\rm M}_\odot$ (similar to ours) can show a deficit of small-scale power of around 30\% at $k\sim0.1$ s/km compared to a higher resolution simulation with 8 times better mass resolution. This difference diminishes with redshift. Given that we are mostly interested in the {\it relative} difference between the CDM and \ethostwo{} models, however, this difference is not critical; the comparison with observations serves mainly as a consistency check of our procedure for generating mock spectra from our simulations.

Bearing in mind that higher resolution only increases the small-scale power by 10-30\%, Fig.~\ref{fig:obsCompare} shows that the \ethostwo{} model is in clear tension with the data on scales smaller than $k\sim0.04$ s/km at $z\lesssim4.6$. This is expected, considering that the linear theory cutoff in the \ethostwo{} model is similar to that of a 1.6 keV thermal relic, which may already be ruled out by existing Lyman-$\alpha$ constraints \citep[e.g.][]{Viel2013,Baur2016}. However, as we have remarked in Section~\ref{sect:intro}, constraining models against observed data by means of their relative normalisation is fraught with uncertainties due to the assumed thermal history of the IGM. We are therefore cautious of our interpretation of Fig.~\ref{fig:obsCompare} with this caveat in mind.

In Fig.~\ref{fig:ratioCompare} we show the redshift evolution of the ratio of the (mean) flux spectra. This figure reveals the defining characteristics of the \ethostwo{} model. At $z\leq4.2$, the behaviour relative to CDM is similar to what is observed in the case of WDM-like models: agreement with CDM on large-scales\footnote{The increased power on large-scales in \ethostwo{} compared to CDM is simply an artefact of rescaling the mean flux. As the small-scale power is suppressed heavily in the \ethostwo{} simulation, the large-scale power is boosted somewhat in order to achieve the same mean flux in the two models.}, followed by largely suppressed power below some characteristic scale. At $z\geq5$, however, a `bump' develops at $k\sim0.13\rmn{~s~km}^{-1}$, which becomes increasingly prominent at higher redshift. This is, indeed, the imprint of the DAO in the gas distribution at these early times. This feature is even more prominent at $z=6$, where even the {\it second} DAO is visible at $k\sim0.2$ s/km. In contrast, the `bump' at $k\sim0.4 \rmn{~s~km}^{-1}$ that becomes increasingly prominent towards high redshifts is most likely a numerical effect associated with the finite resolution in our simulations, which affects different cosmologies differently. As the overall power increases across all scales between $z=6$ to $z=5.4$, mode coupling due to the (mildly) non-linear evolution erases the second DAO bump and transfers its power to smaller and larger scales. As a result, the first DAO peak moves towards smaller scales. Note that the flux spectrum  at $z=6$ is shown simply for comparison, and does not necessarily represent the true ratio at this redshift, where the rescaling procedure may no longer be valid due to incomplete reionisation (see the discussion in Section~\ref{sect:lya}).

Fig.~\ref{fig:ratioCompare} reveals the value of Lyman-$\alpha$ flux spectrum as a probe of small-scale clustering: while the 3D DM distribution showed no evidence of DAOs at $z\lesssim10$ (Fig.~\ref{fig:nlPk}), the linear scales probed by the flux power spectrum bears memory of the acoustic oscillations in the linear power spectrum of the \ethostwo{} model. This may be because the 1D flux spectrum, which can be qualitatively understood as an integrated version of the 3D power spectrum along the line-of-sight, weighted by velocity moments, is more sensitive to small-scale features in the linear power spectrum than the 3D clustering. This is somewhat reminiscent of modified theories of gravity (e.g. the $f(R)$ gravity model), in which the velocity divergence power spectrum (an integral of motion) has been shown to be a much more sensitive probe of deviations from standard gravity than simply the matter density field \citep[e.g.][]{Jennings2012,Bose2015}. We leave a full understanding of the comparison between 1D and 3D power spectra for future work.

\begin{figure}
\centering
\includegraphics[width=\columnwidth]{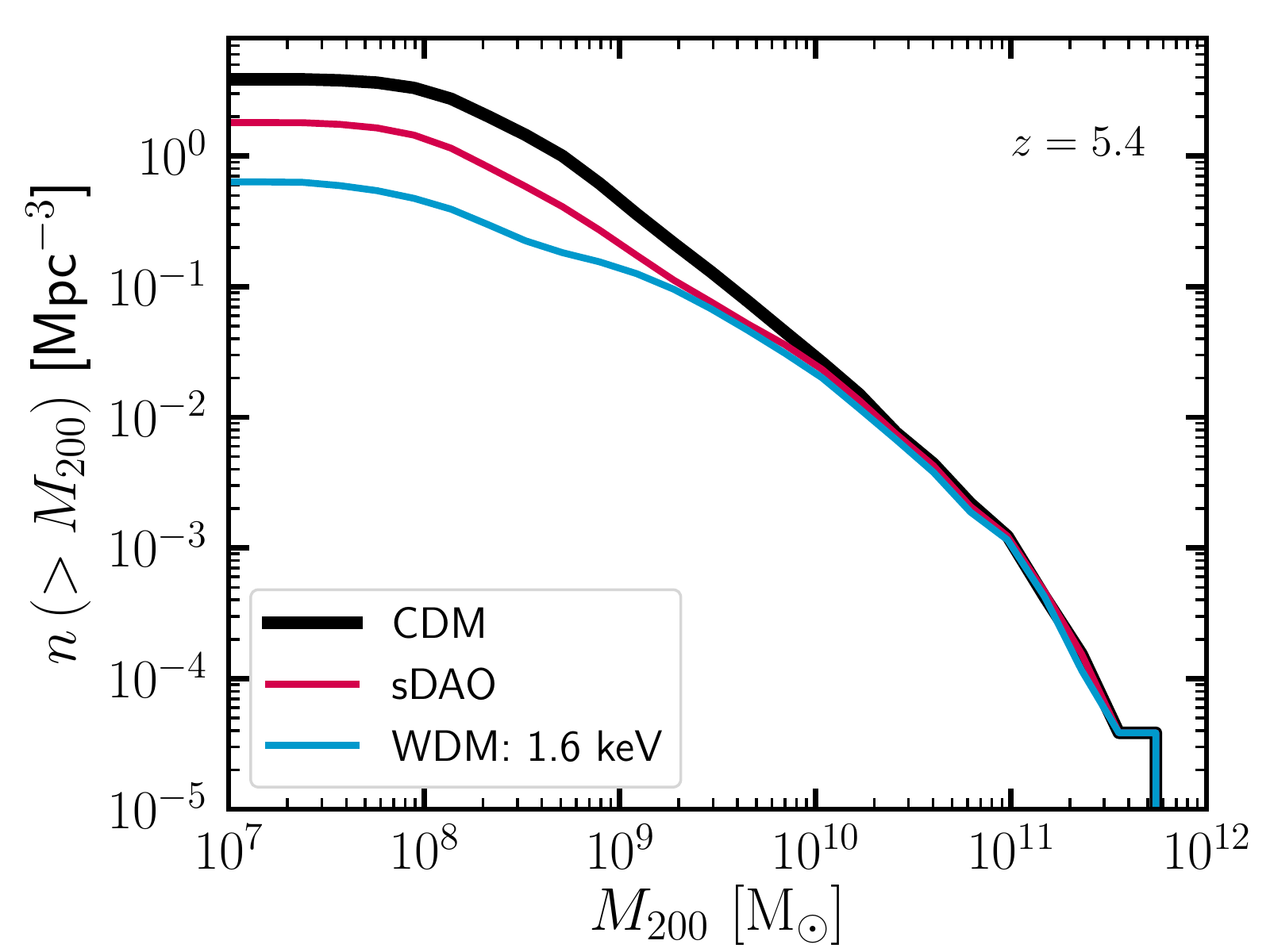}
\caption{Comparison of the cumulative halo mass functions at $z=5.4$ for the CDM, \ethostwo{} and 1.6 keV thermal relic models. While both the \ethostwo{} and WDM models begin to deviate from CDM at a similar mass scale, there is more small-scale power in \ethostwo{}. The noticeable upturn at $M_{200}\sim3\times10^8\,{\rm M}_\odot$ in the WDM mass function is the tell-tale signature of artificial fragmentation \citep{Wang2007}; this is largely absent in the \ethostwo{} model.}
\label{fig:hmfCompareModels}
\end{figure}

It is illuminating to consider the difference in structure in the \ethostwo{} and WDM models at these early times in greater detail. Fig.~\ref{fig:hmfCompareModels} compares the (cumulative) halo mass function in CDM, \ethostwo{} and 1.6 keV cosmologies at $z=5.4$. In this calculation, halo mass is defined by $M_{200}$, which is the mass contained within $r_{200}$, the radius interior to which the mean density is equal to 200 times the critical density of the universe at that redshift. As expected, all three models agree on the abundance of the most massive haloes in the volume at these times ($M_{200} > 10^{10}\,{\rm M}_\odot$). Both the \ethostwo{} and WDM models then peel-away from the CDM curve at an identical mass scale; this is a direct consequence of the fact that the linear power spectra of these two models also deviate from CDM at identical scales. There is, however, a clear excess (of around a factor of 3) of haloes with $M_{200}<3\times10^9\,{\rm M}_\odot$ in \ethostwo{} compared to the 1.6 keV simulation. This excess of power is sourced by the DAO, whereas the initial density fluctuations are suppressed indefinitely in the case of WDM. It is also interesting to note that while the effects of artificial halo formation is clear in the WDM case (as evidenced by the unnatural `upturn' in the mass function at $M_{200}\sim3\times10^8\,{\rm M}_\odot$; \citealt{Wang2007}), the manifestation of these spurious haloes seems largely reduced in the \ethostwo{} model, in which any spurious halo formation is outnumbered by haloes that have collapsed out of true gravitational instability.

\begin{figure*}
\centering
\includegraphics[width=0.49\textwidth]{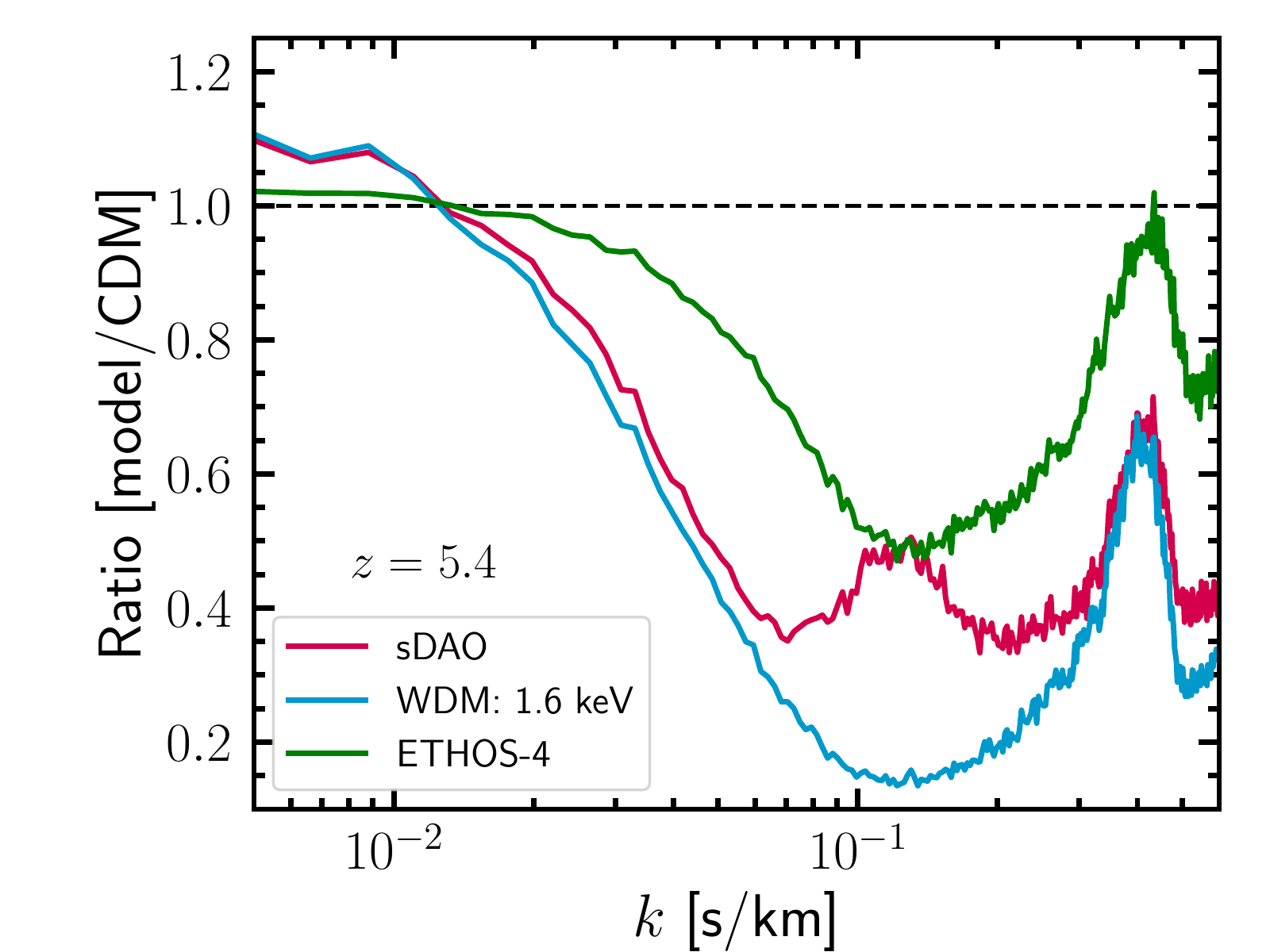}
\includegraphics[width=0.49\textwidth]{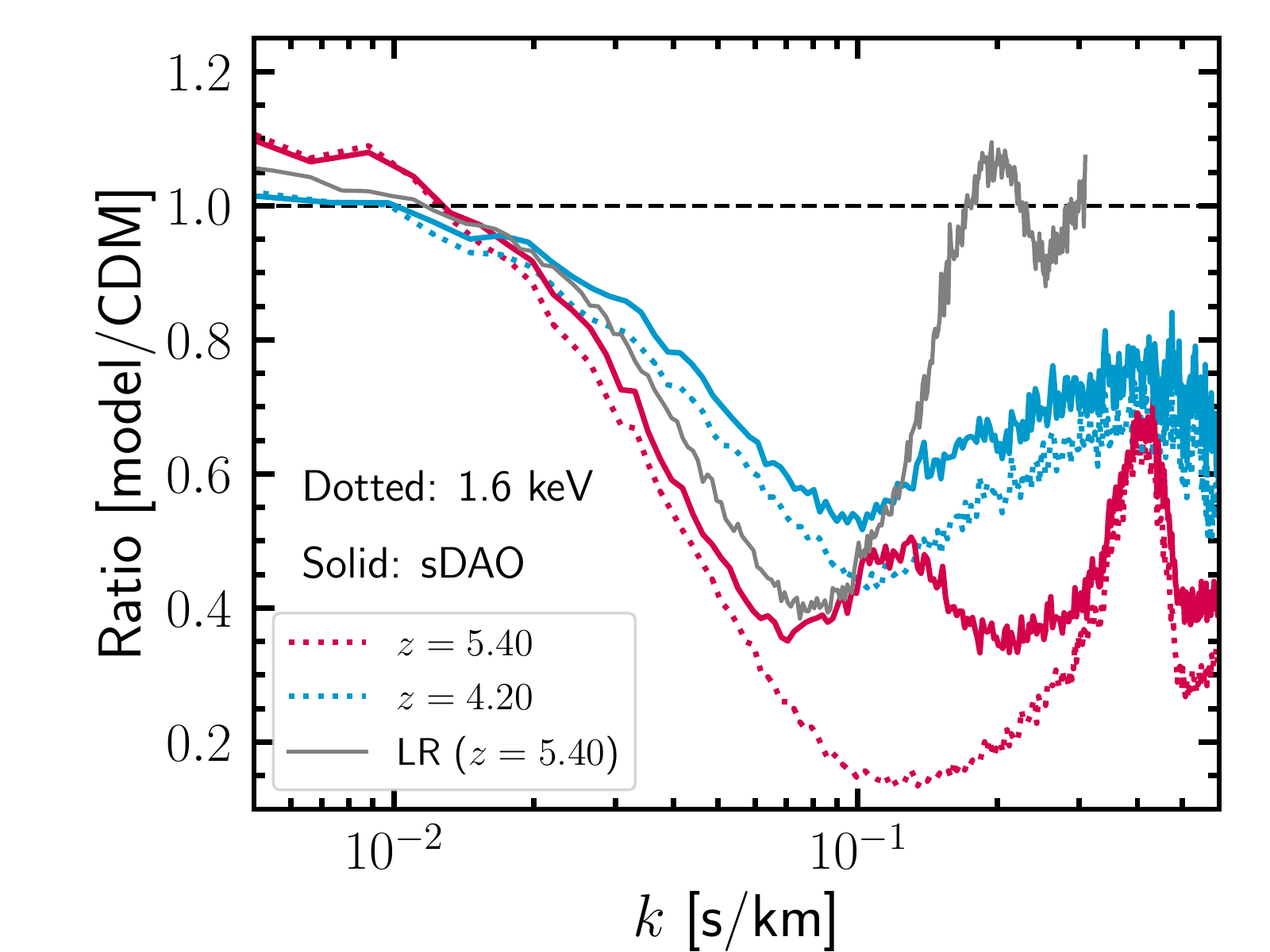}
\caption{{\it Left panel}: As Fig.~\ref{fig:ratioCompare}, but now comparing the \ethostwo{} model with the 1.6 keV WDM and \ethosfour{} models at $z=5.4$. The qualitative behaviour of each model is similar on scales larger than $k=0.1$ s/km, in that power is suppressed relative to CDM. On smaller scales, the \ethostwo{} model exhibits a prominent bump induced by the first DAO peak. This feature is not observed in \ethosfour{}, which also exhibits DAOs in the linear power spectrum, but of smaller amplitude than in the \ethostwo{} case. Each model has been run at the same resolution and each curve therefore exhibits the same numerical `bump' at $k=0.4$ s/km. {\it Right panel}: The evolution of the flux spectrum ratio from $z=5.4$ to $z=4.2$ for the \ethostwo{} (solid lines) and 1.6 keV WDM models (dotted lines). The $z=5.4$ flux spectrum ratio in the low-resolution (LR) \ethostwo{} simulation is shown in grey. As time progresses, the overall increase of power causes adjacent modes to couple non-linearly, thus erasing any sharp (DAO) features in the power spectrum, until $z=4.2$ where the behaviour of the \ethostwo{} and 1.6 keV WDM models is very similar.}
\label{fig:PkCompareModels}
\end{figure*}

The left panel of Fig.~\ref{fig:PkCompareModels} compares the relative difference of the flux spectra to CDM in the two models at $z=5.4$. Power on scales larger than $k\sim0.05$ s/km is suppressed by an almost identical amount, but the behaviour of the two models is different on smaller scales. In particular, while power continues to be suppressed in the case of the 1.6 keV thermal relic, the cutoff in the power is halted by the development of the DAO bump around $k\sim0.13\rmn{~s~km}^{-1}$, which is only present in the sDAO model and not in the WDM model. In practice, this may prove to be difficult to observe since the largest signal is expected to be present at the highest redshift, where the UV background starts to be inhomogeneous due to incomplete reionisation.

We also show predictions for the \ethosfour{} model in which the cutoff is on a smaller scale than in the \ethostwo{} case, and where the first DAO peak is of lower amplitude 
than in \ethostwo{} and is pushed to smaller scales (see Fig.~\ref{fig:inputPk}). The DAO feature in \ethosfour{} is thus unresolved by our simulation (the numerical setup was selected to just resolve the first \ethostwo{} peak). Regardless, this comparison highlights the potential of 1D flux spectrum measurements to distinguish not only non-CDM models from CDM, but also different non-CDM models from each other. The major constraining power comes from scales smaller than $k\sim0.08$ s/km, where there is only limited data available at the moment \citep[but see][for newer data reaching to somewhat smaller scales]{Boera2018}.

One may be concerned that the DAO features we have identified in the $z=5.4$ flux power spectrum may be affected by the small-scale noise manifest as the artificial peak at $k\sim0.4$ s/km. To diagnose this, in the right panel of Fig.~\ref{fig:PkCompareModels} we show the evolution of the flux power spectrum ratio from $z=5.4$ to $z=4.2$ for the \ethostwo{} and 1.6 keV WDM models. At $z=5.4$, the DAO is very prominently present in the \ethostwo{} case while it is of course absent for the 1.6 keV model; on the other hand, the behaviour of the two models is almost identical by $z=4.2$. This is consistent with the picture in Fig.~\ref{fig:ratioCompare}: the second DAO, which was visible at $z=6$, is smoothed away by $z=5.4$ due to non-linear mode coupling; similarly, the first DAO bump, which is visible at $z=5.4$, is smeared away by $z=4.2$. This is because the overall power across all scales increases towards lower redshift, giving the illusion of the DAO peaks being smeared with the numerical ``noise peak'' as time progresses. The effects of noise in the flux power spectrum are manifest more strongly in the 1.6 keV WDM case as there is a lack of ``real'' power on small-scales, in contrast to the \ethostwo{} model where the acoustic oscillation adds physical power on a level larger than the noise at $k>0.1$ s/km.

How the noise level shifts as a function of resolution (see also \citealt{Viel2013}) may be evaluated by comparing the $z=5.4$ flux spectra for the sDAO model at low and high resolution (grey and red curves, respectively). At the lower resolution, the numerical bump is shifted to larger scales by a factor of two (as expected, since the low-resolution simulation retains the same number of particles in a box that is twice as big as the high-resolution simulation). Moreover, the DAO bump, which just starts to develop, blends with the numerical bump and is therefore unresolved in the low-resolution simulation. With increased resolution (i.e., in our default simulations), the DAO is resolved before the noise becomes dominant. Thus, this figure reassures us that our physical interpretation of the first peak in the $z=5.4$ flux spectrum for the \ethostwo{} model is not affected strongly by numerical systematics. As in the case of the cutoff in the small-scale flux spectrum, it may be that quantitative details in Fig.~\ref{fig:PkCompareModels} are affected by assumptions made for the thermal history of the IGM. While varying these assumptions may certainly smear the prominence of the DAO feature, it is not clear that such bumps could be replicated by baryonic mechanisms. In particular, the scale at which these features are manifest, if induced by the nature of the DM, will be set by processes intrinsic to the DM model. We leave the detailed investigation of degeneracies between DAOs and thermal histories to future work.

\section{Conclusions}
\label{sect:conclusions}
We have performed detailed hydrodynamical simulations of non-standard dark matter (DM) species in which the DM is coupled to a relativistic component in the early universe. These interactions alter the primordial linear power spectrum predicted by the concordance cosmological model in a distinctive way: by generating a cutoff at the scale of dwarf galaxies through collisional damping, followed subsequently by a series of `dark acoustic oscillations' (DAOs) towards smaller scales (see Fig.~\ref{fig:inputPk}). Early structure formation in these models is therefore modified considerably from standard cold dark matter (CDM), principally in the form of a delay in the formation of the first stars, and a suppression in the abundance of low-mass galaxies \citep[e.g.][]{Lovell:2017eec}. The structure of DM haloes may be modified as well through strong DM self-interactions at late-times that reshape the phase-space density profiles of galactic haloes \citep[e.g.][]{Vogelsberger2016}. The extent to which these processes impact galaxy formation are, of course, sensitive to parameters specific to the DM theory, such as the duration of DM-radiation coupling, or the self-interaction cross-section. 

While it is impossible to explore this parameter space fully, various permutations of these model parameters will predict largely similar galactic populations. The \ethos{} framework \citep{CyrRacine2016} provides a formalism for mapping these DM properties to `effective' parameters that shape structure formation, thereby providing a flexible way to explore the implications of a vast range of theories on galaxy formation. In this paper, we focus our attention on an atomic DM model (which we refer to as \ethostwo{}) in which DM is composed of two massive fermions that are oppositely charged under a new unbroken $U(1)$ dark gauge force (see Section~\ref{sect:ethos_desc}). The linear matter power spectrum of this model has a cutoff relative to CDM at $k\sim10\,h$cMpc$^{-1}$, identical to a warm dark matter (WDM) thermal relic with mass 1.6 keV, but differs from WDM on smaller scales where it is composed of a significant number of undamped DAOs. While models as extreme as these may already be strongly constrained, our goal in this paper was to investigate if DAOs may be, in principle, detectable in the Lyman-alpha forest, rather than to present a model that matches the available data. {\it A priori}, it is not obvious that DAOs would persist in the Lyman-alpha
flux spectrum. In particular, we sought to identify observational proxies that are able to distinguish between the different small-scale behaviour of these DAO models from WDM. For this purpose, we have investigated the statistics of the Lyman-$\alpha$ forest extracted from hydrodynamical simulations performed with these models using the \arepo{} code \citep{Springel2010} coupled with a sophisticated galaxy formation model used as part of the IllustrisTNG project \citep{Marinacci2018, Naiman2018, Nelson2018, Pillepich2018b, Springel2018}.

Our main conclusions from the current study are:
\begin{enumerate}
    \item On scales smaller than $k\sim4\,h$ cMpc$^{-1}$, the 3D distribution of DM is clustered less strongly in the \ethostwo{} model than in CDM, although the differences get smaller with time (Fig.~\ref{fig:nlPk}). In particular, while there is a strong DAO signature imprinted in the matter distribution at $z\geq10$, further epochs of gravitational collapse wash away this feature entirely by $z\sim6$.
    \item A random line-of-sight through the \ethostwo{} simulation box reveals far less structure in absorption than the equivalent line-of-sight in the CDM simulation (Fig.~\ref{fig:los_spectrum}). This is a direct consequence of the cutoff in the primordial power spectrum in the \ethostwo{} model, delaying the formation of galaxies at these high redshifts ($z\gtrsim3)$.  
    \item Despite the delayed start to the galaxy formation process in the \ethostwo{} model, it catches up with CDM by $z\approx2$. This faster growth of structure is a fairly generic phenomenon observed in models with a cutoff in the linear power spectrum (including WDM). In our work, this is manifest in the form of the transmitted flux PDFs (Fig.~\ref{fig:fluxPDF}), which are truncated towards high values in the \ethostwo{} model at $z\geq4$, but are identical to CDM by $z=2$. The probability that a given line-of-sight intersects a region with high transmitted flux increases as the universe transitions from neutral to ionised due to the ionising radiation from high redshift galaxies.
    \item While the 1D flux power spectra are identical in CDM and in the \ethostwo{} model at $z\leq3$, there are significant differences at higher redshift. In fact, present data at these redshifts already place the \ethostwo{} in significant tension with observations (Fig.~\ref{fig:obsCompare}), although astrophysical systematics may relax the level of discrepancy.
    \item More interestingly, however, we find that the DAO bump characteristic of the \ethostwo{} model -- which was absent in the 3D matter distribution -- is imprinted prominently in the 1D flux power spectrum at $z\geq5$ on scales smaller than $k\sim0.1$ s/km (Fig.~\ref{fig:ratioCompare}). At $z\leq4.2$, the DAO feature is smoothed out, and the behaviour of the model is then reminiscent of standard WDM. 
    \item The appearance and disappearance of the DAO at different redshifts therefore offers an opportunity to disentangle small-scale features in the flux power spectrum induced by the nature of DM from astrophysical effects (e.g. different reionisation histories). In particular, precise measurements of the flux power spectrum on scales smaller than $k\approx0.1$ s/km will be fundamental to distinguishing different DM models from each other (Fig.~\ref{fig:PkCompareModels}). 
\end{enumerate}

While there is a vast parameter space of well-motivated non-standard CDM models, the predictions they make for the formation of structure and the properties of galaxies can be challenging to differentiate. Of fundamental importance is the need to identify sets of statistics that allow the identification of {\it physical} scales that are characteristic of these theories. DM models in which there is a coupling to a relativistic species in the early Universe are characterised in the linear regime by a cutoff at the scale of dwarf galaxies followed by a series of dark acoustic oscillations towards smaller scales. In this work we have shown that these fundamental scales, while absent in the total matter distribution after the epoch of reionisation, are imprinted in the 1D Lyman-$\alpha$ flux power spectrum in a way that may be constrained with future high-precision observations. In the meantime, it is interesting to consider further statistics that could reveal the scale-dependent behaviour of different DM theories; possible examples include the clustering of DLAs and Lyman-limit systems or cross-correlations of Lyman-$\alpha$ with galaxy properties.


\appendix
\section{Resolution tests}
\begin{figure}
\centering
\includegraphics[width=\columnwidth]{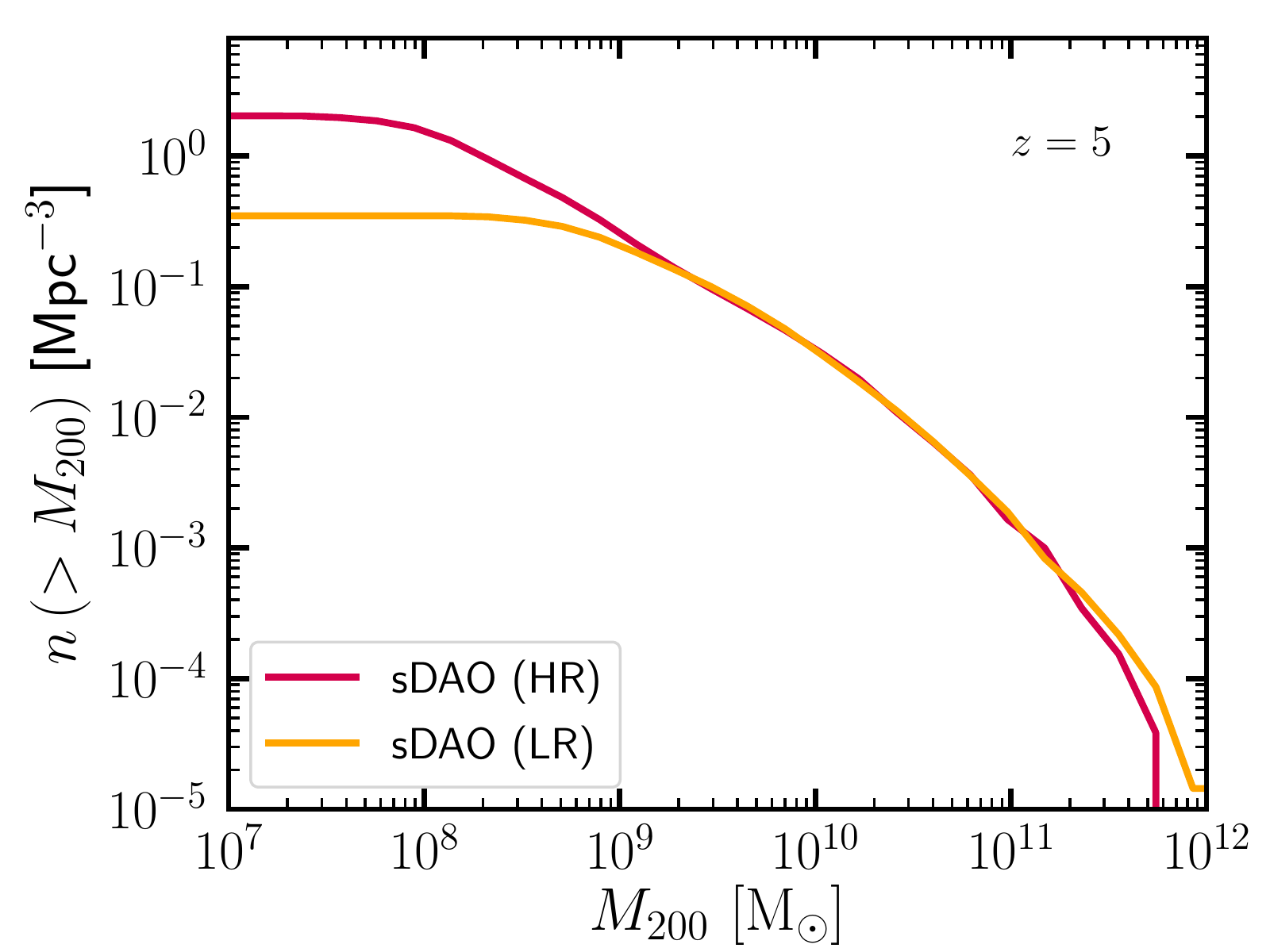}
\caption{Comparison of the cumulative halo mass functions at $z=5$ in the \ethostwo{} LR (40 $h^{-1}\,{\rm Mpc}$, $N_p=2\times512^3$) and HR (20 $h^{-1}\,{\rm Mpc}$, $N_p=2\times512^3$) simulations. The excess at the high mass end in the LR simulation is due to the larger box size; similarly, the lack of low mass haloes is due to the lower resolution than in the HR version. Note that the characteristic upturn due to spurious fragmentation, which is clearly visible in WDM models \citep{Wang2007} is no longer clear in the sDAO model at either resolution.}
\label{fig:hmfCompare}
\end{figure}

In Fig.~\ref{fig:hmfCompare}, we compare the $z=5$ halo mass functions measured in the \ethostwo{} high-resolution (HR, 20 $h^{-1}\,{\rm Mpc}$, $N_p=2\times512^3$) and \ethostwo{} low-resolution (LR, 40 $h^{-1}\,{\rm Mpc}$, $N_p=2\times512^3$) simulations. The mass functions are converged over the expected range. The LR simulation shows an excess at the very massive end, as these rare haloes are more likely to be found in the larger volume of the LR simulation. It is interesting note, however, that neither simulation shows an upturn towards the low mass end, which is usually the characteristic signature of artificial halo formation in WDM simulations. This problem is exacerbated at low resolution -- the scale below which these fragments start to dominate scales with the number of particles roughly as $N_p^{1/3}$ \citep{Wang2007}. While this scale is clearly present in the 1.6 keV WDM simulation (see Fig.~\ref{fig:hmfCompareModels}), the effect is largely suppressed in the \ethostwo{} model at both LR and HR. The reason for this is that while the cutoff in WDM continues indefinitely, the DAO adds power on scales smaller than the initial cutoff, thereby largely offsetting the instability through discreteness effects that is typical of WDM-like simulations.

\begin{figure}
\centering
\includegraphics[width=\columnwidth]{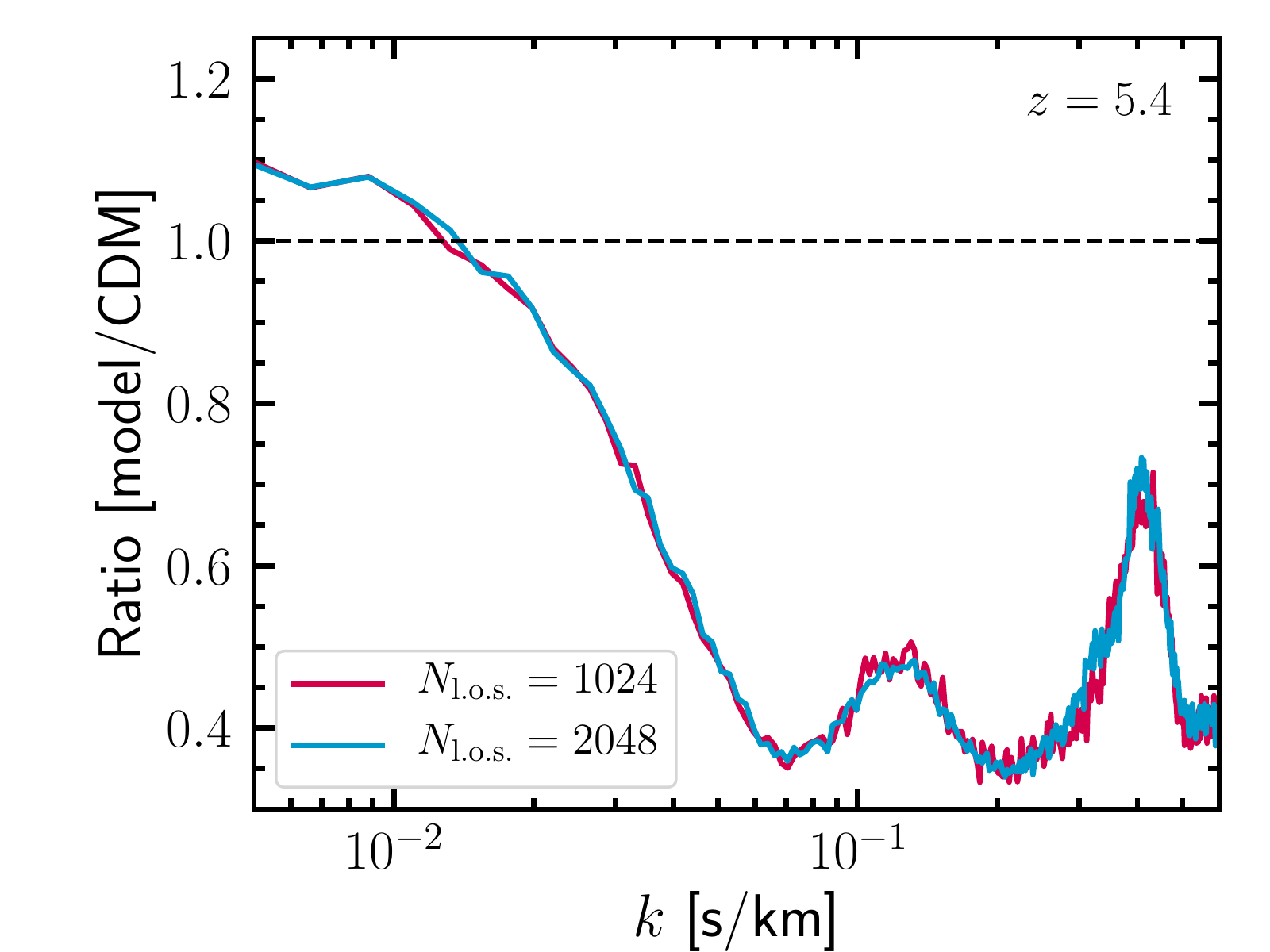}
\caption{Demonstrating the convergence of the 1D Lyman-$\alpha$ flux spectrum ratio with respect to the number of lines-of-sight chosen for the analysis.}
\label{fig:NlosCompare}
\end{figure}

Fig.~\ref{fig:NlosCompare} justifies our use of 1024 lines-of-sight for generating mock absorption spectra from our simulations. In this figure we show the ratio (\ethostwo{}/CDM) of the 1D flux power spectrum at $z=5.4$ using 1024 (red) and 2048 (blue) lines-of-sight. The two curves show excellent convergence across all scales. We have checked explicitly that individual power spectra (rather than simply the ratio) are converged as well. We find that, in general, the truncated Gaussian smoothing kernel \citep[][see also Eq.~\ref{eq:trunc}]{Altay2013} is relatively robust to the number of skewers used to generate the mock spectra.

\section{Dark matter model parameters}\label{app:model_params}
We list in Table \ref{tab:sDAO_model_params} the atomic DM parameters used to generate our sDAO model. For more details on the model, see \cite{Cyr-Racine:2013ab}.   

 \begin{table}
     \centering
     \begin{tabular}{c|c|c}
    \hline \hline
         Parameter & Description & Value   \\
         \hline
         $\xi$ &  $T_{{\rm DR}}/T_{{\rm CMB}}$  &0.15 \\
         $\alpha_D$ & The dark fine structure constant & 0.02  \\
         $B_D$ & The dark atom binding energy & 1.7 keV  \\
         $m_{{\rm DM}}$ & The dark atom mass & 500 MeV \\
         \hline
     \end{tabular}
     \caption{Atomic DM particle parameters used to generate our sDAO model. $T_{\rm DR}$ is the temperature of the dark radiation bath, while $T_{\rm CMB}$ is the cosmic microwave background temperature.}
     \label{tab:sDAO_model_params}
 \end{table}
 
 
\section*{Acknowledgements}
We thank the anonymous referee for providing  suggestions that have improved this manuscript. We are very grateful to Volker Springel for allowing us access to \arepo{}, which was used to run all the simulations used in this paper. SB is supported by Harvard University through the ITC Fellowship. MV acknowledges support through an MIT RSC award, a Kavli Research Investment Fund, NASA ATP grant NNX17AG29G, and NSF grants AST-1814053 and AST-1814259. JZ and Sebastian Bohr acknowledge support by a Grant of Excellence from the Icelandic Research Fund (grant number 173929$-$051). CP acknowledges support by the European Research Council under ERC-CoG grant CRAGSMAN-646955. F-Y C-R acknowledges the support of the National Aeronautical and Space Administration ATP grant NNX16AI12G at Harvard University. This work was made possible in part by usage of computing resources at the University of Southern Denmark through the NeIC Dellingr resource sharing pilot. A fraction of the simulations in this work were carried out on the Garpur supercomputer, a joint project between the University of Iceland and University of Reykjav\'ik with funding from Rann\'is.

\bibliographystyle{mnras} \bibliography{refs.bib}{}

\bsp	
\label{lastpage}
\end{document}